\documentclass[a4paper,12pt]{article}
\usepackage[a4paper,text={16.8cm,22.4cm}]{geometry}
\usepackage{amsmath,amssymb,latexsym,bm,braket}
\usepackage{cite}
\usepackage{graphicx}

\allowdisplaybreaks 
\newcommand{\Tr}{\mathrm{Tr}}

\begin{document}

\begin{titlepage}

  \begin{flushright}
    MZ-TH/12-29 \\
    ZU-TH 13/12 \\
    July 20, 2012
  \end{flushright}
  
  \vspace{5ex}
  
  \begin{center}
    \textbf{ \Large The NNLO soft function for the pair invariant mass
      distribution of boosted top quarks} \vspace{7ex}
    
    \textsc{Andrea Ferroglia$^a$, Ben D. Pecjak$^b$, and Li Lin
      Yang$^c$} \vspace{2ex}
  
    \textsl{${}^a$New York City College of Technology, 300 Jay Street\\
      Brooklyn, NY 11201, USA
      \\[0.3cm]
      ${}^b$Institut f\"ur Physik (THEP), Johannes Gutenberg-Universit\"at\\
      D-55099 Mainz, Germany
      \\[0.3cm]
      ${}^c$Institute for Theoretical Physics, University of Z\"urich\\
      CH-8057 Z\"urich, Switzerland}
  \end{center}

  \vspace{4ex}

  \begin{abstract}
    At high values of the pair invariant mass the differential cross
    section for top-quark pair production at hadron colliders
    factorizes into soft, hard, and fragmentation functions. In this
    paper we calculate the next-to-next-to-leading-order (NNLO)
    corrections to the soft function appearing in this factorization
    formula, thus providing the final piece needed to evaluate at NNLO
    the differential cross section in the virtual plus soft
    approximation in the large invariant-mass limit. Technically, this
    amounts to evaluating the vacuum expectation value of a soft
    Wilson loop operator built out of light-like Wilson lines for each
    of the four partons participating in the hard scattering process,
    with a certain constraint on the total energy of the soft
    radiation. Our result turns out to be surprisingly simple, because
    in the sum of all graphs the three and four parton contributions
    multiply color structures whose coefficients are governed by the
    non-abelian exponentiation theorem.
  \end{abstract}

\end{titlepage}

\section{Introduction}
\label{sec:intro}

The pair invariant mass distribution is an important observable in
top-quark pair production at hadron colliders. Especially interesting
is the high invariant-mass region of the distribution, which could
potentially be distorted by new physics without disturbing the good
agreement between Standard Model and experiment for the total cross
section. The phenomenological importance of this distribution
motivates special attention to its calculation in the Standard Model.

Theory predictions of the invariant mass distribution typically rely
on the next-to-leading order (NLO) computations of two-particle
inclusive cross sections carried out in \cite{Mangano:1991jk},
supplemented with soft gluon resummation at the level of
next-to-leading-logarithmic (NLL) \cite{Kidonakis:1997gm,
  Almeida:2008ug} or more recently next-to-next-to-leading-logarithmic
(NNLL) \cite{Ahrens:2010zv} accuracy. The resummed calculations take
into account higher-order logarithmic plus-distribution corrections
related to gluon emission in the soft limit $z=M^2/\hat{s}\to 1$, with
$M$ the pair invariant mass and $\sqrt{\hat{s}}$ the partonic
center-of-mass energy. However, except at restrictively high values of
the invariant mass where $\tau=M^2/s \to 1$, with $\sqrt{s}$ the
hadronic center-of-mass energy, the dominance of these logarithmic
corrections over $\delta(1-z)$ corrections relies on the mechanism of
dynamical threshold enhancement studied in \cite{Appell:1988ie,
  Catani:1998tm, Becher:2007ty, Bauer:2010jv}. In fact, numerical
comparisons of exact NLO results with the leading terms in the soft
limit show good agreement only if the delta-function terms are
included to achieve a full virtual plus soft approximation
\cite{Ahrens:2009uz}. Starting at NNLO, such a virtual plus soft
approximation is not achieved through expansions of NNLL resummation
formulas alone, and even though the delta-function contributions are
formally N$^3$LL corrections the only way to know their size for
certain is to calculate them.\footnote{The delta-function terms and
  also non-singular terms in the soft limit would be included in a
  full NNLO calculation, but current NNLO results are limited to
  certain $q\bar q$-initiated contributions to the total cross section
  \cite{Baernreuther:2012ws,Czakon:2012zr}.} Moreover, the soft gluon
resummation mentioned above uses the generic counting $m_t\sim M$ for
the top-quark mass, whereas at truly high values of the invariant mass
the counting $m_t\ll M$ should be used, and resummation must also take
into account logarithms of the ratio $m_t/M$.

In a recent paper \cite{Ferroglia:2012ku}, we have set up a
factorization formalism appropriate for the invariant-mass
distribution in the simultaneous soft and small-mass ($m_t\ll M$)
limit. Schematically, the factorization is of the form
\begin{align}
  \label{eq:cartoonfact}
  d\hat{\sigma} = \Tr \left[\bm{H}\bm{S}\right]\otimes D\otimes D
  +{\cal O}(1-z)+{\cal O}\left(\frac{m_t}{M}\right)\,.
\end{align}
The hard function $\bm{H}$ and the soft function $\bm{S}$ are matrices
in the space of color-singlet operators for $(q\bar q,\,gg) \to t\bar
t$ scattering, evaluated with $m_t=0$, while the $D$ are perturbative
heavy-quark fragmentation functions containing the dependence on
$m_t$. Given this form of factorization, one can derive and solve
renormalization-group (RG) equations for the individual functions to
resum soft logarithms as well as those depending on $m_t/M$. Just as
importantly, the individual functions appearing in
(\ref{eq:cartoonfact}) are much easier to calculate in fixed-order
perturbation theory than the hard and soft functions needed for soft
gluon resummation for generic $m_t$. In fact, the fragmentation
function is completely known to NNLO accuracy \cite{Melnikov:2004bm},
and the higher-order virtual corrections to two-to-two scattering
needed to extract the contribution of the NNLO hard function to the
differential cross section are also known \cite{Anastasiou:2000kg,
  Anastasiou:2000mv, Glover:2001af, Glover:2001rd, Anastasiou:2001sv}.
The only missing piece needed to obtain at NNLO a full soft plus
virtual approximation in the limit of large invariant mass is the soft
function $\bm{S}$. The calculation of this soft function to NNLO is
the subject of this paper.

Our main motivation for this calculation is its eventual impact on the
phenomenology of top-quark pair production. However, we consider it an
interesting problem even apart from this. The soft function we deal
with here is related to double real emission corrections to massless
two-to-two scattering, and at the technical level is defined as the
vacuum expectation value of a Wilson loop built out of four light-like
Wilson lines. While a number of soft functions have been calculated at
NNLO in the literature, these all involve either two
\cite{Belitsky:1998tc, Becher:2005pd, Kelley:2011ng, Monni:2011gb,
  Hornig:2011iu, Li:2011zp, Kelley:2011aa} or three
\cite{Becher:2012za} Wilson lines. Compared to those cases, the
four-Wilson-line soft function is characterized by the added
conceptual complication of a non-trivial matrix structure in color
space. One might also expect it to be computationally much more
complicated because, unlike the case studied in \cite{Becher:2012za},
graphs with attachments of gluons to three Wilson lines do not vanish
and in general are complicated functions of two non-trivial scalar
products. However, we find that in the sum of all diagrams such
three-parton contributions multiply a color structure whose
coefficient is determined by the non-abelian exponentiation theorem
\cite{Gatheral:1983cz, Frenkel:1984pz}. In particular, the bare
function does not contain a three parton term with the antisymmetric
color structure of a three-gluon vertex. This is an expected result
for the IR divergent pieces of the bare function, as a consequence of
the form of the RG equation derived in \cite{Ferroglia:2012ku}. In
particular, the momentum dependence in the anomalous dimension
governing this RG equation is inherited from the anomalous dimension
for scattering amplitudes and is of the dipole type at least to NNLO,
which follows from results in \cite{Aybat:2006wq, Aybat:2006mz} (and
may even be true to all orders, as conjectured in \cite{Becher:2009cu,
  Gardi:2009qi, Becher:2009qa}). For the IR finite pieces it is
perhaps slightly unexpected that the three parton terms are determined
by non-abelian exponentiation, but this is nonetheless the case due to
cancellations among certain diagrams.

The remainder of this paper is organized as follows. First, in
Section~\ref{sec:definitions} we give the precise definition of the
soft function calculated in this paper and we also review the NLO
calculation from \cite{Ferroglia:2012ku}, using it as a means of
introducing some of the formalism related to the color-space matrix
structure. Then, in Section~\ref{sec:NNLObare} we present an
expression for the bare NNLO soft function as a sum over legs and give
explicit results for the component integrals and matrix structures
appearing in this sum. We also describe cross checks, both with the
two-Wilson-line integrals calculated in \cite{Li:2011zp} and with the
non-abelian exponentiation theorem. Finally, in
Section~\ref{sec:renormalization}, we discuss the renormalization
procedure and explain how this provides a further cross check on our
result. We conclude in Section~\ref{sec:conclusions}, relegating
several details of the calculation along with the final results for
the NNLO soft function to the appendix.

\section{Definitions and the NLO calculation}
\label{sec:definitions}

We define the soft function for the pair-invariant mass distribution
as in \cite{Ahrens:2010zv}, adapted to the massless case by replacing
time-like velocity vectors by light-like ones \cite{Ferroglia:2012ku}.
The basic objects for the soft function are firstly the Wilson lines
\begin{align}
  \label{eq:Wilson}
  \bm{S}_i(x) = \mathcal{P} \exp \left( ig_s \int_{-\infty}^{0} ds \;
    n_i\cdot A^a(x+sn_i) \; \bm{T}_i^a \right) ,
\end{align}
where the $\mathcal{P}$ refers to path ordering, and secondly the
Wilson loop built out of these objects,
\begin{align}
  \label{eq:WilsonLoop}
  \bm{O}_s(x) = \big[ \bm{S}_{n_1} \bm{S}_{n_2} \bm{S}_{n_3}
  \bm{S}_{n_4} \big] (x) \, .
\end{align}
We have used the notation of \cite{Catani:1996jh, Catani:1996vz},
where the bold-face indicates that the objects ${\bm T}_i^a$ are
matrices acting on color structures specific to the type of partons
participating in the two-to-two scattering process. This notation
allows us to describe simultaneously the case where the Wilson lines
$\bm{S}_i$ associated with the four partons are in any combination of
the fundamental (for quarks) or adjoint (for gluons) representations.
In this paper we have in mind applications to top-quark pair
production in the soft limit, and so will give results appropriate for
$(q\bar q,gg)\to t\bar t$ scattering.

The Wilson-loop operator (\ref{eq:WilsonLoop}) takes into account the
coupling of soft gluons to the external partons within the eikonal
approximation. The soft function is related to the contribution of
these gluon emissions at the level of the squared amplitude. In
\cite{Ferroglia:2012ku}, it was defined through the Fourier transform
of a position-space soft function evaluated in the parton
center-of-mass frame. In the present work, we will find it more
convenient to work directly with the following momentum-space
representation\footnote{Note that this definition differs from
  \cite{Ferroglia:2012ku} by a factor of $\sqrt{\hat s}$, which we
  have chosen to omit here. The Laplace-transformed function in
  (\ref{eq:stilde}), on the other hand, coincides with the one
  introduced in \cite{Ferroglia:2012ku}.}:
\begin{align}
  \label{eq:Sdef}
  \bm{S}(\omega,t_1/M^2,\mu) = \frac{1}{d_R} \sum_{X_s}
  \braket{0|\bm{O}_s^\dagger(0)|X_s} \braket{X_s|\bm{O}_s(0)|0} \,
  \delta(\omega - (n_1+n_2)\cdot p_{X_s}) \, ,
\end{align}
where $d_R=N$ in the $q\bar q$ channel, and $d_R = N^2-1$ in the $gg$
channel, with $N$ the number of colors. As usual, $X_s$ refers to a
final state built up of any number of unobserved soft gluons. It is
clear from (\ref{eq:Sdef}) that the soft function depends on the
single dimensionful parameter $\omega$ (and $\mu$, upon
renormalization), as well as the scalar products $n_i\cdot n_j$.
Although the results we give later on can be used to construct the
soft function for arbitrary velocity vectors, we have defined it in
the natural way for two-to-two scattering. In that case there are only
two independent scalar products and thus one non-trivial dimensionless
ratio, which we have chosen as $n_1\cdot n_3/n_1\cdot n_2= n_2\cdot
n_4/n_1\cdot n_2= -t_1/M^2$. Our notation is then in direct
correspondence with the Mandelstam variables used in
\cite{Ferroglia:2012ku}. It further implies that $n_1\cdot
n_4/n_1\cdot n_2 =n_2\cdot n_3/n_1\cdot n_2= 1+t_1/M^2 \equiv
-u_1/M^2$ and $n_1\cdot n_2=n_3\cdot n_4=2$. In the parton
center-of-mass-frame, the delta function constrains the energy of the
soft radiation to $2E_s=\omega$, and it is an easy matter to show the
correspondence with the position-space definition used in
\cite{Ferroglia:2012ku}.

In order to study the higher-order corrections which are the subject
of this paper, we define expansion coefficients of the bare soft
function in $d=4-2\epsilon$ dimensions as
\begin{align}
  \label{eq:Sexp}
  \bm{S}_{\rm bare} = & \bm{S}^{(0)} +
  \left(\frac{Z_{\alpha}\alpha_s}{4\pi}\right)\bm{S}_{\rm bare}^{(1)}
  + \left(\frac{Z_{\alpha}\alpha_s}{4\pi}\right)^2 \bm{S}_{\rm
    bare}^{(2)}+ \cdots \, .
\end{align} 
In the above equation we have expressed the bare coupling constant
$\alpha_s^{(0)}$ in terms of the renormalized one in the
$\overline{\text{MS}}$ scheme: the relation between the two is
$Z_\alpha \alpha_s \mu^{2\epsilon}=e^{- \epsilon \gamma_E}
(4\pi)^{\epsilon}\alpha_s^{(0)}$ with $Z_\alpha=1-\beta_0
\alpha_s/(4\pi\epsilon)$ and $\beta_0 = 11/3 N -2/3 n_l$, with $n_l$
the number of light flavors. The soft function for massless partons
was obtained to NLO in \cite{Ferroglia:2012ku}. We end this section by
reviewing the elements that go into that calculation. This gives us an
opportunity to introduce the aspects of the color-space formalism
needed in this work, and to extend our previous results to the depth
of the $\epsilon$-expansion needed for the NNLO calculation.

The color-space formalism provides a means of representing soft gluon
emissions from external quarks and gluons in a unified way. The
important point for the case at hand is that these soft emissions can
mix the possible color-singlet structures appearing in the two-to-two
scattering amplitudes. To explain the matrix structure relevant for
this mixing, we must first define a color basis for $(q^{a_1}\bar
q^{a_2},g^{a_1}g^{a_2})\to t^{a_3}\bar t^{a_4}$ scattering, where the
$\{a\}$ label the color indices of the partons with velocity $n_i$. We
work in the $s$-channel singlet-octet basis
\begin{gather}
  \label{eq:colorstructures}
  \big( c^{q\bar{q}}_1 \big)_{\{a\}} = \delta_{a_1a_2} \delta_{a_3a_4}
  \, , \qquad \big( c^{q\bar{q}}_2 \big)_{\{a\}} = t^c_{a_2a_1}
  t^c_{a_3a_4} \, , \nonumber
  \\
  \big( c^{gg}_1 \big)_{\{a\}} = \delta^{a_1a_2} \delta_{a_3a_4} \, ,
  \qquad \big( c^{gg}_2 \big)_{\{a\}} = if^{a_1a_2c} \, t^c_{a_3a_4}
  \, , \qquad \big( c^{gg}_3 \big)_{\{a\}} = d^{a_1a_2c} \,
  t^c_{a_3a_4} \, .
\end{gather}
We view these structures as basis vectors $\ket{c_I}$ in the space of
color-singlet amplitudes. Inner products in this space are defined
through a summation over color indices as
\begin{align}
  \label{eq:innerproducts}
  \Braket{c_I|c_J}= \sum_{\{a\}}\big(c_I\big)^*_{a_1 a_2 a_3 a_4}
  \big(c_J\big)_{a_1 a_2 a_3 a_4} \,.
\end{align}
This inner product is proportional but not equal to $\delta_{IJ}$, so
the basis vectors are orthogonal but not orthonormal. The soft
function matrix elements are defined as
\begin{align}
  \label{eq:Smatdef}
  \bm{S}_{IJ}=\bra{c_I}\bm{S}\ket{c_J}\,.
\end{align} 
The soft function is thus a two-by-two matrix in the $q\bar q$
channel, and a three-by-three matrix in the $gg$ channel. At leading
order (LO), a short calculation yields the result
\begin{align}
  \label{eq:Stree}
  \bm{S}_{q\bar{q}}^{(0)} = \delta(\omega)
  \begin{pmatrix}
    N & 0
    \\
    0 & \frac{C_F}{2}
  \end{pmatrix}
  , \qquad \bm{S}_{gg}^{(0)} = \delta(\omega)
  \begin{pmatrix}
    N & 0 & 0
    \\
    0 & \frac{N}{2} & 0
    \\
    0 & 0 & \frac{N^2-4}{2N}
  \end{pmatrix}
  \,.
\end{align}
At NLO and beyond gluon emissions from eikonal lines are associated
with factors of $\bm{T}_i$. One determines the matrix elements
(\ref{eq:Smatdef}) by using a set of rules which dictate how the
$\bm{T}_i$ act on the color indices $\{a\}$ of the color structures
(\ref{eq:colorstructures}) and performing the sum over colors
indicated in (\ref{eq:innerproducts}). These rules are as follows
\cite{Catani:1996jh, Catani:1996vz}: if the $i$-th parton is a
final-state quark or an initial-state anti-quark we set
$(\bm{T}^c_i)_{ba} = t^c_{ba}$, for a final-state anti-quark or an
initial-state quark we have $(\bm{T}^c_i)_{ba} = -t^c_{ab}$, and for a
gluon we use $(\bm{T}^c_i)_{ba} = if^{abc}$. In (\ref{eq:Sdef}) there
is also a conjugated operator $\bm{O}_s^\dagger$, which contains
conjugated color matrices $\bm{T}^{a\dagger}_i$ acting to the left.
However, we can drop the daggers and let them act to the right, using
the relation
\begin{align}
  \braket{\mathcal{M} | \bm{T}^{a\dagger}_i | \mathcal{M'}} = \left(
    \bm{T}^a_i \ket{\mathcal{M}} \right)^\dagger \ket{\mathcal{M}'} =
  \ket{\mathcal{M}}^\dagger \left( \bm{T}^a_i \ket{\mathcal{M}'}
  \right) = \braket{\mathcal{M} | \bm{T}^{a}_i | \mathcal{M'}} \, ,
\end{align}
with $\ket{\mathcal{M}}$ and $\ket{\mathcal{M}'}$ being two arbitrary
vectors in color space. This can be easily understood by noting that
$\bm{T}^a_i$ is a Hermitian operator. To show explicitly that this
relation holds, it is sufficient to work out the case where $i=1$ and
$\ket{\mathcal{M}}$ and $\ket{\mathcal{M}'}$ are two basis vectors
$\ket{a_1,\ldots}$ and $\ket{b_1,\ldots}$. We then have
\begin{align*}
  \left( \bm{T}^c_1 \ket{a_1,\ldots} \right)^\dagger \ket{b_1,\ldots}
  = \left( (\bm{T}^c_1)_{d_1a_1} \ket{d_1,\ldots} \right)^\dagger
  \ket{b_1,\ldots} = (\bm{T}^c_1)^*_{d_1a_1} \delta_{b_1d_1} =
  (\bm{T}^c_1)_{a_1b_1} \, ,
  \\
  \ket{a_1,\ldots}^\dagger \left( \bm{T}^c_1 \ket{b_1,\ldots} \right)
  = \ket{a_1,\ldots}^\dagger \left( (\bm{T}^c_1)_{d_1b_1}
    \ket{d_1,\ldots} \right) = (\bm{T}^c_1)_{d_1b_1} \delta_{a_1d_1} =
  (\bm{T}^c_1)_{a_1b_1} \, .
\end{align*}
Note also that if there is more than one $\bm{T}$ matrix acting on the
same Wilson line, the conjugation reverses their order, e.g.,
\begin{align}
  \left( \bm{T}^a_i \, \bm{T}^b_i \ket{\mathcal{M}} \right)^\dagger
  \ket{\mathcal{M}'} = \ket{\mathcal{M}}^\dagger \left( \bm{T}^b_i \,
    \bm{T}^a_i \ket{\mathcal{M}'} \right) .
\end{align}
This is crucial for the cancellations of certain contributions which
we will encounter later on.

\begin{figure}
  \begin{center}
    \includegraphics{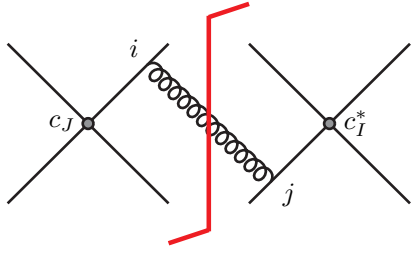}
  \end{center}
  \caption{Diagram contributing to the NLO soft
    function. \label{fig:NLOsoft}}
\end{figure}

The formalism is now in place to discuss the NLO calculation. This
requires us to evaluate the type of diagram shown in
Figure~\ref{fig:NLOsoft} and perform a sum over attachments to the
different legs. The diagram represents the connection of gluons to
Wilson lines for two distinct particles $i$ and $j$. Diagrams
involving virtual corrections or two attachments to the same line are
scaleless and vanish in dimensional regularization. To evaluate the
diagram we associate each gluon emission with the Feynman rules for
eikonal attachments following from (\ref{eq:Wilson}) and set the cut
propagator (for which we use Feynman gauge) on shell with positive
energy. This results in an expression proportional to the integral
\begin{align}
  \label{eq:I1}
  I_1(\omega,a_{ij}) &= \int [dk] \, \frac{n_i \cdot n_j \;
    \delta(\omega - n_0 \cdot k)}{n_i \cdot k \; n_j \cdot k} \equiv
  \pi^{1-\epsilon} \, e^{-\epsilon\gamma_E} \; \omega^{-1-2\epsilon}
  \; \bar{I}_1(a_{ij}) \, ,
\end{align}
where $[dk]= d^dk \, \delta(k^2) \, \theta(k^0)$ and
\begin{align}
  a_{ij} \equiv 1 - \frac{n_0^2 \; n_i \cdot n_j}{2 \, n_0 \cdot n_i
    \; n_0 \cdot n_j} \, .
\end{align}
For top-quark production, where $n_0^2=(n_1+n_2)^2=4$, we have
$a_{12}=a_{34}=0$, $a_{13}=a_{24}=-u_1/M^2$, $a_{14}=a_{23}=-t_1/M^2$.
Later on we will describe cross-checks of our result with the simpler
NNLO soft functions for Drell-Yan \cite{Belitsky:1998tc} and
electroweak boson production at large $p_T$ \cite{Becher:2012za}.
These involve the same two-Wilson-line integrals but with $a=0$ and
$a=1$ respectively.

The integral in (\ref{eq:I1}) can be parameterized in terms of the
gluon energy and an angular integral which is of the type considered
long ago in \cite{vanNeerven:1985xr}. We have rederived the result
using the light-cone coordinate decomposition from
\cite{Becher:2012za}, which turns out to be especially convenient for
the NNLO integrals considered later on. Either way, the result for the
stripped integral is
\begin{align}
  \bar{I}_1(a) &= \frac{2 \, e^{\epsilon\gamma_E} \,
    \Gamma(-\epsilon)}{\Gamma(1-2\epsilon)} \, (1-a)^{-\epsilon} \;
  {}_2F_1(-\epsilon,-\epsilon,1-\epsilon,a) \,,\label{eq:I1res}
\end{align}
which can be expanded in $\epsilon$ using
\begin{align}
  {}_2F_1(-\epsilon,-\epsilon,1-\epsilon,a) &= 1 + H_{2}(a) \epsilon^2
  + \left( H_{3}(a) - H_{2,1}(a) \right) \epsilon^3 \nonumber
  \\
  &+ \left( H_{4}(a) - H_{3,1}(a) + H_{2,1,1}(a) \right) \epsilon^4 +
  \cdots \,. \label{eq:HG1}
\end{align}
The functions $H$ indicate Harmonic Polylogarithms\footnote{We write
  the HPLs in the compact notation which eliminates the zeros in the
  weight vector by adding at the same time one to the absolute value
  of the previous index to the right; for example $H_{2}(a) =
  H(0,1;a)$, $H_{4}(a) = H(0,0,0,1;a)$, $H_{2,1,1}(a) = H(0,1,1,1;a)$,
  etc.} (HPLs) \cite{Remiddi:1999ew}. Here and elsewhere in the paper
we have used the Mathematica package HypExp \cite{Huber:2005yg} in
expanding the hypergeometric functions and manipulated the resulting
HPLs with the package HPL \cite{Maitre:2005uu}.

We obtain the bare NLO soft function through the following sum over legs:
\begin{align}
  \bm{S}^{(1)}_{\rm bare} &= \frac{2}{\omega}
  \left(\frac{\mu}{\omega}\right)^{2\epsilon} \, \sum_{\text{legs}} \,
  \bm{w}_{ij}^{(1)} \, \bar{I}_1(a_{ij}) \,\nonumber
  \\
  &= \frac{4}{\omega} \left(\frac{\mu}{\omega}\right)^{2\epsilon}
  \left( \bm{w}_{12}^{(1)} \, \bar{I}_1(a_{12})+\bm{w}_{34}^{(1)} \,
    \bar{I}_1(a_{12})+2 \bm{w}_{13}^{(1)} \, \bar{I}_1(a_{13}) +2
    \bm{w}_{14}^{(1)} \, \bar{I}_1(a_{14})\right)\, .
\end{align}
We have taken into account the relations between the $a_{ij}$ and also
the explicit form of the color matrices given below to simplify the
sum. The matrix structure in color space is obtained by evaluating the
matrix elements of
\begin{align}
  \bm{w}^{(1)}_{ij} &= -\frac{1}{d_R} \, \bm{T}_i \cdot \bm{T}_j \, ,
\end{align}
as in (\ref{eq:Smatdef}). Results were given in \cite{Ahrens:2010zv},
and we reprint them here for convenience. In the $q\bar q$ channel
they read
\begin{align}
  \label{eq:qTT}
  \bm{w}_{12}^{(1)} = \bm{w}_{34}^{(1)} &= \frac{C_F}{4N}
  \begin{pmatrix}
    4N^2 & 0
    \\
    0 & -1
  \end{pmatrix}
  , \nonumber
  \\
  \bm{w}_{13}^{(1)} = \bm{w}_{24}^{(1)} &= \frac{C_F}{2}
  \begin{pmatrix}
    0 & 1
    \\
    1 & 2C_F - \frac{N}{2}
  \end{pmatrix}
  , \nonumber
  \\
  \bm{w}_{14}^{(1)} = \bm{w}_{23}^{(1)} &= \frac{C_F}{2N}
  \begin{pmatrix}
    0 & -N
    \\
    -N & 1
  \end{pmatrix}
  ,
\end{align}
while for the $gg$ channel they are 
\begin{align}
  \label{eq:gTT}
  \bm{w}_{12}^{(1)} &= \frac{1}{4}
  \begin{pmatrix}
    4N^2 & 0 & 0
    \\
    0 & N^2 & 0
    \\
    0 & 0 & N^2-4
  \end{pmatrix}
  , \nonumber
  \\
  \bm{w}_{34}^{(1)}&=
  \begin{pmatrix}
    C_FN & 0 & 0
    \\
    0 & -\frac{1}{4} & 0
    \\
    0 & 0 & -\frac{N^2-4}{4N^2}
  \end{pmatrix}
  , \nonumber
    \\
 \bm{w}_{13}^{(1)}= \bm{w}_{24}^{(1)}&= \frac{1}{8}
  \begin{pmatrix}
    0 & 4N & 0
    \\
    4N & N^2 & N^2-4
    \\
    0 & N^2-4 & N^2-4
  \end{pmatrix}
  , \nonumber
  \\
  \bm{w}_{14}^{(1)}= \bm{w}_{23}^{(1)}&= \frac{1}{8}
  \begin{pmatrix}
    0 & -4N & 0
    \\
    -4N & N^2 & -(N^2-4)
    \\
    0 & -(N^2-4) & N^2-4
  \end{pmatrix}
  .
\end{align}
As usual, $C_F=(N^2-1)/2N$ and $C_A=N$.  

At NLO the renormalized function in the $\overline{\text{MS}}$ scheme
can be obtained from the bare function simply by dropping the poles.
More formally, we need to multiply it on both sides by a UV
renormalization matrix. We describe the formal procedure in more
detail in Section~\ref{sec:renormalization}, after performing the NNLO
calculation in the next section.

\section{The bare soft function at NNLO}
\label{sec:NNLObare}

In this section we calculate the bare soft function at NNLO. We find
that in the sum of all diagrams the result can be written in the form
\begin{align}
  \label{eq:S2bare}
  \bm{S}^{(2)}_{\text{bare}} &= \frac{4}{\omega} \left(
    \frac{\mu}{\omega} \right)^{4\epsilon} \sum_{\text{legs}} \left(
    \sum_{n=2}^{7} \bm{w}^{(n)}_{ij} \, \bar{I}_n(a_{ij}) +
    \bm{w}^{(8)}_{ijk} \, \bar{I}_8(a_{ij},a_{ik}) +
    \bm{w}^{(9)}_{ijkl} \, \bar{I}_9(a_{ij},a_{kl}) \right)
  \\
  &= \frac{4}{\omega} \left( \frac{\mu}{\omega} \right)^{4\epsilon}
  \left[ 2 \left( \bm{w}^{(1)}_{12} + \bm{w}^{(1)}_{34} \right) \left(
      \bar{I}_2(a_{12}) + C_A\bar{I}_6(a_{12}) + C_A
      \bar{I}_{7,1}(a_{12}) + C_A \bar{I}_{7,2}(a_{12}) \right)
  \right. \nonumber
  \\
  &+ 4\bm{w}^{(1)}_{13} \left( \bar{I}_2(a_{13}) + C_A
    \bar{I}_6(a_{13}) + C_A \bar{I}_{7,1}(a_{13}) + C_A
    \bar{I}_{7,2}(a_{13}) \right) \nonumber
  \\
  &+ 4\bm{w}^{(1)}_{14} \left( \bar{I}_2(a_{14}) + C_A
    \bar{I}_6(a_{14}) + C_A \bar{I}_{7,1}(a_{14}) + C_A
    \bar{I}_{7,2}(a_{14}) \right) \nonumber
  \\
  &+ 2 \left( \bm{w}^{(3)}_{12} + \bm{w}^{(3)}_{34} \right)
  \bar{I}_3(a_{12}) + 4 \bm{w}^{(3)}_{13} \left( \bar{I}_3(a_{13}) +
    \bar{I}_4(a_{13}) \right) + 4 \bm{w}^{(3)}_{14} \left(
    \bar{I}_3(a_{14}) + \bar{I}_4(a_{14}) \right) \nonumber
  \\
  &+ 4 \bm{w}^{(9)}_{12} \, \bar{I}_4(a_{12}) + \left(
    \bm{w}^{(4)}_{12} + \bm{w}^{(4)}_{34} \right) \left(
    \bar{I}_4(a_{12}) + 2\bar{I}_5(a_{12}) \right) \nonumber
  \\
  &+ 2 \bm{w}^{(4)}_{13} \left( \bar{I}_4(a_{13}) + 2\bar{I}_5(a_{13})
  \right) + 2 \bm{w}^{(4)}_{14} \left( \bar{I}_4(a_{14}) +
    2\bar{I}_5(a_{14}) \right) \label{eq:Legs}
  \\
  &+ 4 \left( \bm{w}^{(8)}_{123} + \bm{w}^{(8)}_{314} \right)
  \bar{I}_8(a_{12},a_{13}) + 4 \left( \bm{w}^{(8)}_{124} +
    \bm{w}^{(8)}_{324} \right) \bar{I}_8(a_{12},a_{14}) + 8
  \bm{w}^{(8)}_{134} \, \bar{I}_8(a_{13},a_{14}) \, . \nonumber
\end{align}
Three types of basic diagrams contribute to this sum, depending on
whether the gluons attach to two, three, or four distinct Wilson
lines. We organize this section by discussing each type of diagram in
term, and give explicit results for the color factors and integrals
appearing in (\ref{eq:Legs}). The results for the three and four
parton diagrams turn out to be surprisingly simple, because the
non-abelian exponentiation theorem constrains the coefficients of the
color structures $\bm{w}^{(8)}_{ijk}$ and $\bm{w}^{(9)}_{ijkl}$ in
(\ref{eq:S2bare}). We discuss this further in
Appendix~\ref{sec:NAexp}.

\subsection{Two-Wilson-line integrals}
\label{sec:2W}

\begin{figure}
  \centering
  \includegraphics{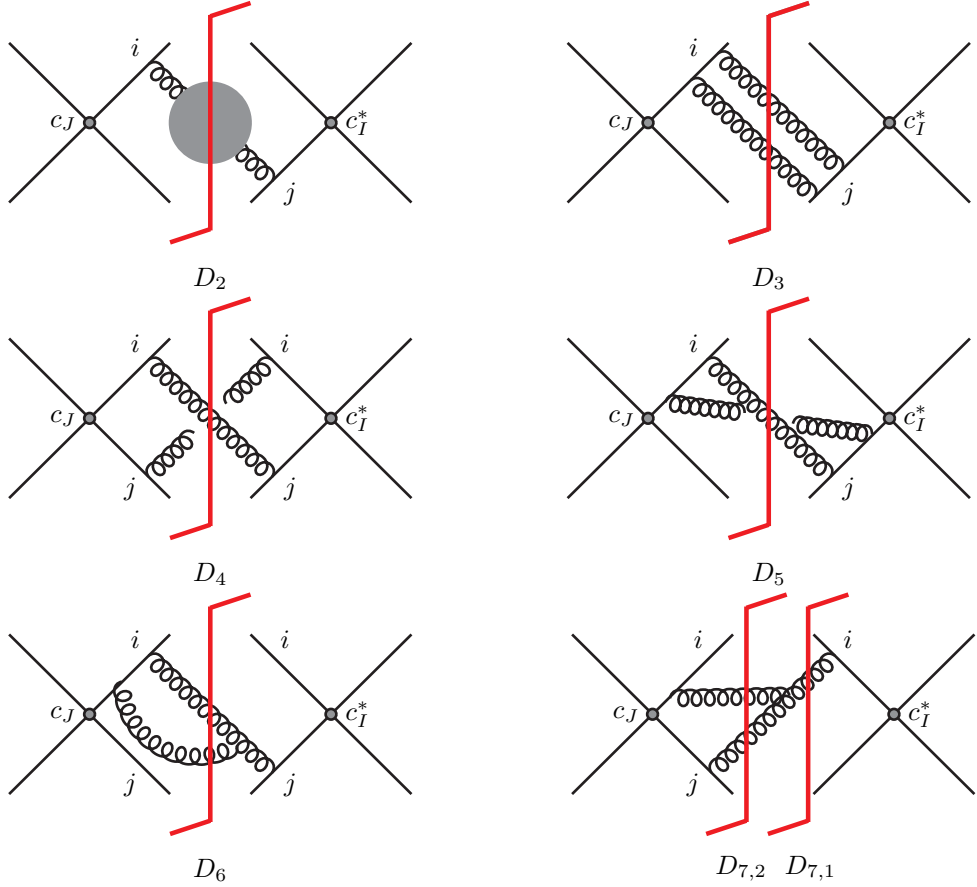}
  \caption{Two-Wilson-line integrals required in the calculation of
    the NNLO soft matrix. \label{fig:NNLO2W}}
\end{figure}

The subset of non-vanishing two-Wilson-line integrals is familiar from
other calculations of soft functions to NNLO \cite{Belitsky:1998tc,
  Li:2011zp, Becher:2012za}. The relevant Feynman diagrams are shown
in Figure~\ref{fig:NNLO2W}. Converting these diagrams into integral
expressions is straightforward, and leads to the following set:
\begin{align}
  \label{eq:Iset}
  I_2(\omega,a_{ij}) &=
  -\pi^{1-\epsilon}\left(C_A(5-3\epsilon)-2N_l(1-\epsilon)\right)
  \frac{\Gamma(2-\epsilon)}{\Gamma(4-2\epsilon)} \int [dk] \,
  \frac{n_i \cdot n_j \; \delta(\omega - n_0 \cdot k)}{n_i \cdot k \;
    n_j \cdot k \; (k^2)^{1+\epsilon}} \, , \nonumber
  \\
  I_3(\omega,a_{ij}) &= \int [dk] \, [dl] \, \frac{(n_i \cdot n_j)^2
    \; \delta(\omega - n_0 \cdot (k+l))}{n_i \cdot k \; n_j \cdot
    (k+l) \; n_j \cdot l \; n_j \cdot (k+l)} \, , \nonumber
  \\
  I_4(\omega,a_{ij}) &= \int [dk] \, [dl] \, \frac{(n_i \cdot n_j)^2
    \; \delta(\omega - n_0 \cdot (k+l))}{n_i \cdot k \; n_i \cdot l \;
    n_j \cdot k \; n_j \cdot l} \, , \nonumber
  \\
  I_5(\omega,a_{ij}) &= \int [dk] \, [dl] \, \frac{(n_i \cdot n_j)^2
    \; \delta(\omega - n_0 \cdot (k+l))}{n_i \cdot k \; n_i \cdot
    (k+l) \; n_j \cdot k \; n_j \cdot (k+l)} \, , \nonumber
  \\
  I_6(\omega,a_{ij}) &= \int [dk] \, [dl] \, \frac{n_i \cdot n_j \;
    n_i \cdot (l-k) \; \delta(\omega - n_0 \cdot (k+l))}{n_i \cdot k
    \; n_i \cdot (k+l) \; n_j \cdot (k+l) \; (k+l)^2} \, , \nonumber
  \\
  I_{7,1}(\omega,a_{ij}) &= -\pi^{1-\epsilon} \,
  \Re[e^{-i\pi\epsilon}] \, \frac{\Gamma^2(1+\epsilon) \,
    \Gamma^3(-\epsilon)}{\Gamma(-2\epsilon)} \int [dq] \left(
    \frac{n_i \cdot n_j}{2n_i \cdot q \; n_j \cdot q}
  \right)^{1+\epsilon} \delta(\omega - n_0 \cdot q) \, , \nonumber
  \\
  I_{7,2}(\omega,a_{ij}) &= \int [dk] \, [dl] \, \frac{n_i \cdot n_j
    \; n_j \cdot (k+2l) \; \delta(\omega - n_0 \cdot (k+l))}{n_i \cdot
    k \; n_j \cdot (k+l) \; n_j \cdot l \; (k+l)^2} \, .
\end{align}
The prefactors in $I_2$ and $I_{7,1}$ arise from the internal loop
integrals, see for instance the discussion in \cite{Belitsky:1998tc,
  Becher:2012za}. To evaluate the phase-space integrals we use
light-cone coordinate techniques and parameterizations described in
the Appendix of \cite{Becher:2012za}. It is then relatively
straightforward to derive results for the integrals in terms of
hypergeometric functions, with the exception of $I_{5}$ and $I_{7,2}$,
which we evaluate as an expansion in $\epsilon$. In
Appendix~\ref{sec:Example} we derive the results for the integral
$I_5$ as an example of the calculational procedure. Defining stripped
integrals according to (with $n>1$)
\begin{align}
  I_n(\omega,a_{ij}) = \pi^{2-2\epsilon} \, e^{-2\epsilon\gamma_E} \;
  \omega^{-1-4\epsilon} \; \bar{I}_n(a_{ij})
\end{align}
the explicit results can be summarized as
\begin{align}
  \bar{I}_4(a) &= \frac{8 e^{2\epsilon\gamma_E} \, \Gamma^2(-\epsilon)
    \, \Gamma(-2\epsilon)}{\Gamma(1-2\epsilon) \, \Gamma(1-4\epsilon)}
  \, (1-a)^{-2\epsilon} \, \big[
  {}_2F_1(-\epsilon,-\epsilon,1-\epsilon,a) \big]^2 \, ,
  \\
  \bar{I}_5(a) &= (1-a)^{-2\epsilon} \left[ -\frac{1}{\epsilon^3} +
    \frac{1}{\epsilon} \left( \frac{7\pi^2}{6} - 4H_2(a) \right) +
    \frac{62}{3} \zeta_3 - 12H_3(a) + 4H_{2,1}(a) \right.
  \\
  &+ \left. \epsilon \left( \frac{\pi^4}{40} + \frac{14\pi^2}{3}
      H_2(a) - 36H_4(a) - 4H_{2,2}(a) + 12H_{3,1}(a) - 4H_{2,1,1}(a)
    \right) + \cdots \right] , \nonumber
  \\
  \bar{I}_3(a) &= \frac{\bar{I}_4(a)}{2} - \bar{I}_5(a) \, ,
  \\
  \bar{I}_6(a) &= \frac{e^{2\epsilon\gamma_E} \, \Gamma^2(-\epsilon)
    \, \Gamma(-2\epsilon)}{\Gamma(2-2\epsilon) \, \Gamma(1-4\epsilon)}
  \, (1-a)^{-2\epsilon} \;
  {}_2F_1(-2\epsilon,-2\epsilon,1-2\epsilon,a) \, ,
  \\
  \bar{I}_2(a) &= - \left( C_A(5-3\epsilon) - 2N_l(1-\epsilon) \right)
  \frac{2 \, \Gamma(2-\epsilon) \,
    \Gamma(2-2\epsilon)}{\Gamma(4-2\epsilon) \, \Gamma(-\epsilon)} \,
  \bar{I}_6(a) \, ,
  \\
  \bar{I}_{7,1}(a) &= - \frac{2 \, e^{2\epsilon\gamma_E} \,
    \Re[e^{-i\pi\epsilon}] \, \Gamma(-2\epsilon) \,
    \Gamma^2(-\epsilon) \, \Gamma^2(1+\epsilon)}{\Gamma(1-4\epsilon)}
  \, (1-a)^{-2\epsilon} \; {}_2F_1(-2\epsilon,-2\epsilon,1-\epsilon,a)
  \, ,
  \\
  \bar{I}_{7,2}(a) &= (1-a)^{-2\epsilon} \left[ -\frac{2}{\epsilon^3}
    + \frac{1}{\epsilon} \left( \frac{5\pi^2}{2} - 6H_2(a) \right) +
    \frac{139}{3} \zeta_3 - 8H_3(a) + 12H_{2,1}(a) \right.
  \\
  &+ \left. \epsilon \left( \frac{\pi^4}{36} + \frac{23\pi^2}{3}
      H_2(a) - 12H_4(a) - 4H_{2,2}(a) + 16H_{3,1}(a) - 24H_{2,1,1}(a)
    \right) + \cdots \right] . \nonumber
\end{align}
To expand these in $\epsilon$ we use (\ref{eq:HG1}) along with
\begin{align}
  {}_2F_1(-2\epsilon,-2\epsilon,1-\epsilon,a) &= 1 + 4H_{2}(a)
  \epsilon^2 + \left( 4H_{3}(a) - 12H_{2,1}(a) \right) \epsilon^3
  \nonumber
  \\
  &\hspace{0.3cm} + \left( 4H_{4}(a) + 4H_{2,2}(a) - 12H_{3,1}(a) + 36
    H_{2,1,1}(a) \right) \epsilon^4 + \cdots \, .
\end{align}
These results can be related to those for a position-space soft
function calculated in \cite{Li:2011zp}, and we have found full
agreement with that work. They furthermore agree with results from
\cite{Belitsky:1998tc} for $a=0$ and from \cite{Becher:2012za} for
$a=1$ as two special cases.
 
We must also evaluate the color factors. For these we find
\begin{align}
  \label{eq:2WT}
  \bm{w}^{(2)}_{ij} &= \bm{w}^{(1)}_{ij} \, , \nonumber
  \\
  \bm{w}^{(3)}_{ij} &= \frac{1}{d_R} \, \bm{T}_i^a \, \bm{T}_i^b \,
  \bm{T}_j^a \, \bm{T}_j^b \, , \nonumber
  \\
  \bm{w}^{(4)}_{ij} &= \bm{w}^{(5)}_{ij} = \frac{1}{d_R} \, \bm{T}_i^a
  \, \bm{T}_i^b \, \bm{T}_j^b \, \bm{T}_j^a =
  \bm{w}^{(3)}_{ij}-\frac{C_A}{2}\bm{w}^{(1)}_{ij} \, , \nonumber
  \\
  \bm{w}^{(6)}_{ij} &= \bm{w}^{(7,2)}_{ij} = \bm{w}^{(7,1)}_{ij}
  =\frac{1}{d_R} \, if^{abc} \, \bm{T}_i^a \, \bm{T}_i^b \, \bm{T}_j^c
  = \, \frac{C_A}{2} \, \bm{w}^{(1)}_{ij} \,.
\end{align}
The relation between $\bm{w}_{ij}^{(3)}$ and $\bm{w}_{ij}^{(4)}$,
along with the result that $I_3=I_4/2-I_5$, ensures that the bare
function satisfies the non-abelian exponentiation theorem. We discuss
this further in Appendix~\ref{sec:NAexp}. Results for the NLO matrices
$\bm{w}^{(1)}_{ij}$ were given in (\ref{eq:qTT}) and (\ref{eq:gTT}).
In the $q\bar q$ annihilation channel the remaining matrices evaluate
to
\begin{align}
  \bm{w}^{(3)}_{12} &= \bm{w}^{(3)}_{34} = \frac{C_F}{2}
  \begin{pmatrix}
    N^2-1 & 0
    \\
    0 & \frac{1}{4N^2}
  \end{pmatrix}
  , \nonumber
  \\
  \bm{w}^{(3)}_{13} &= \bm{w}^{(3)}_{24}  = \frac{C_F}{2}
  \begin{pmatrix}
    1 & \frac{N^2-2}{2N}
    \\
    \frac{N^2-2}{2N} & \frac{N^4-3N^2+3}{4N^2}
  \end{pmatrix}
  , \nonumber
  \\
  \bm{w}^{(3)}_{14} &= \bm{w}^{(3)}_{23} =  \frac{C_F}{2}
  \begin{pmatrix}
    1 & -\frac{1}{N}
    \\
    -\frac{1}{N} & \frac{N^2+3}{4N^2}
  \end{pmatrix}
  ,
\end{align}
whereas in the gluon fusion channel they are 
\begin{align}
  \bm{w}^{(3)}_{12} &= \frac{C_A}{2}
  \begin{pmatrix}
    2 N^2 & 0 & 0
    \\
    0 & \frac{N^2}{4} & 0
    \\
    0 & 0 & \frac{N^2-4}{4}
  \end{pmatrix}
  , \nonumber
  \\
  \bm{w}^{(3)}_{34} &= \frac{C_A}{2}
  \begin{pmatrix}
    \frac{(N^2-1)^2}{2N^2} & 0 & 0
    \\
    0 & \frac{1}{4N^2} & 0
    \\
    0 & 0 & \frac{N^2-4}{4N^4}
  \end{pmatrix}
  , \nonumber
  \\
  \bm{w}^{(3)}_{13} &= \bm{w}^{(3)}_{24} =
  \frac{C_A}{2}
  \begin{pmatrix}
    1 & \frac{N}{4} & \frac{N^2-4}{4N}
    \\
    \frac{N}{4} & \frac{N^2+2}{8} & \frac{N^2-4}{8}
    \\
    \frac{N^2-4}{4N} & \frac{N^2-4}{8} & \frac{(N^2-2)(N^2-4)}{8N^2}
  \end{pmatrix}
  , \nonumber
  \\
  \bm{w}^{(3)}_{14} &= \bm{w}^{(3)}_{23}  =
  \frac{C_A}{2}
  \begin{pmatrix}
    1 & -\frac{N}{4} & \frac{N^2-4}{4N}
    \\
    -\frac{N}{4} & \frac{N^2+2}{8} & -\frac{N^2-4}{8}
    \\
    \frac{N^2-4}{4N} & -\frac{N^2-4}{8} & \frac{(N^2-2)(N^2-4)}{8N^2}
  \end{pmatrix}
  .
\end{align}

\subsection{Three-Wilson-line integrals}
\label{sec:3W}

\begin{figure}
  \centering
  \includegraphics{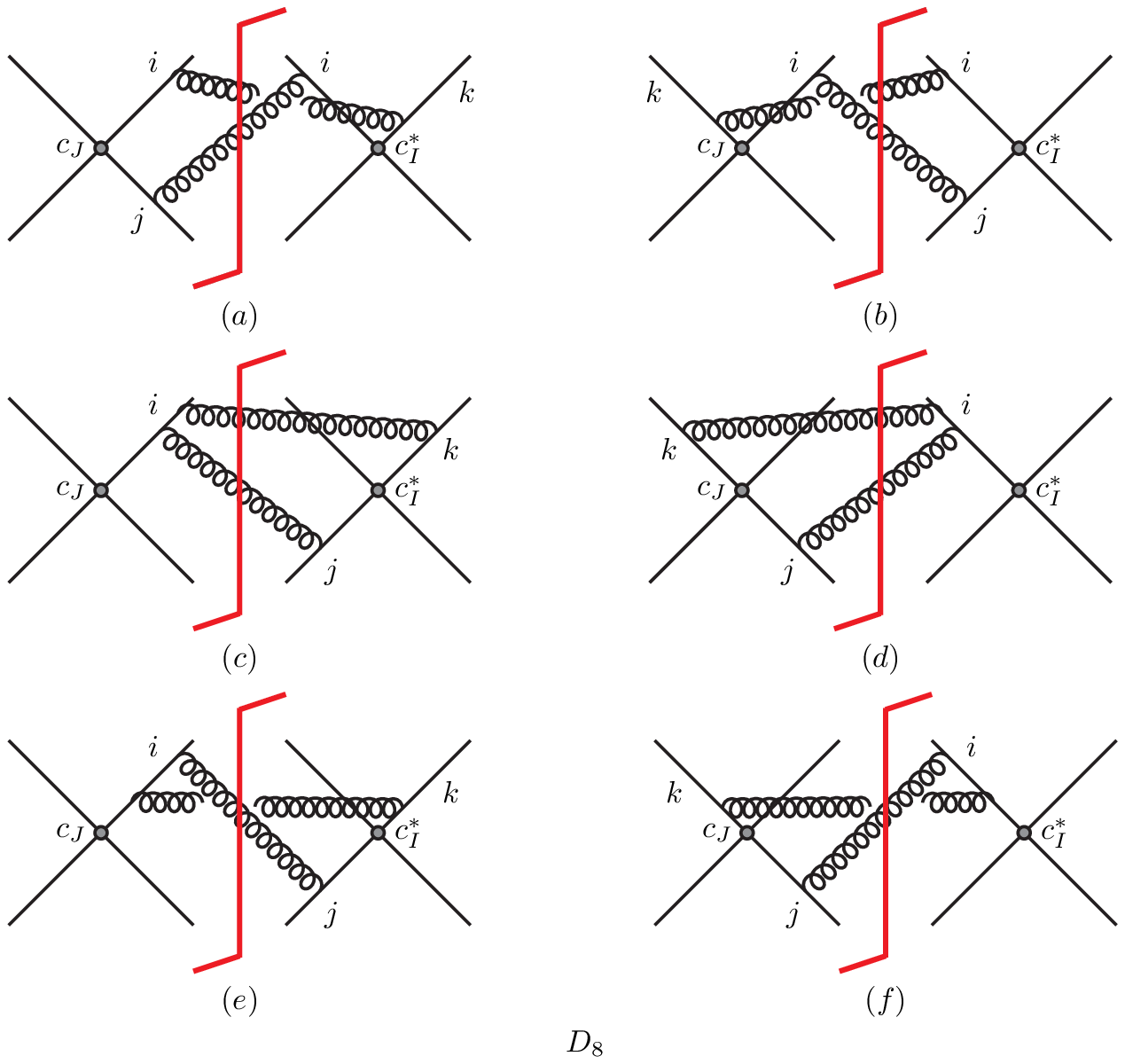}
  \caption{The abelian three-Wilson-line integrals required in the
    calculation of the NNLO soft matrix.\label{fig:NNLO3Wplanar}}
\end{figure}

\begin{figure}[ht]
  \centering
  \includegraphics{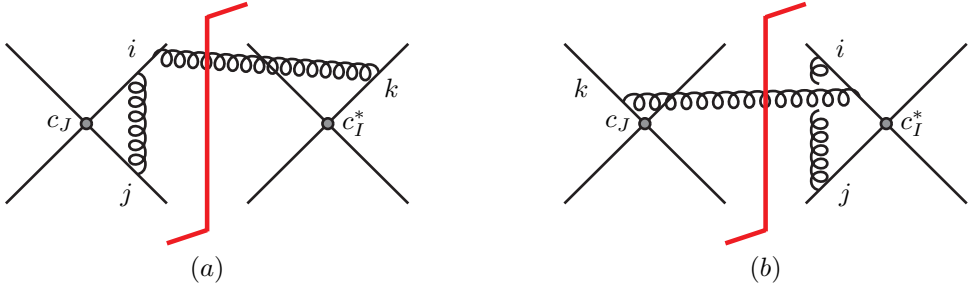}
  \caption{Example of a pair of mixed virtual-real one-particle cuts which
  adds up to a scaleless integral.\label{fig:NNLO3W1LV1LR}}
\end{figure}

\begin{figure}[ht]
  \centering
  \includegraphics{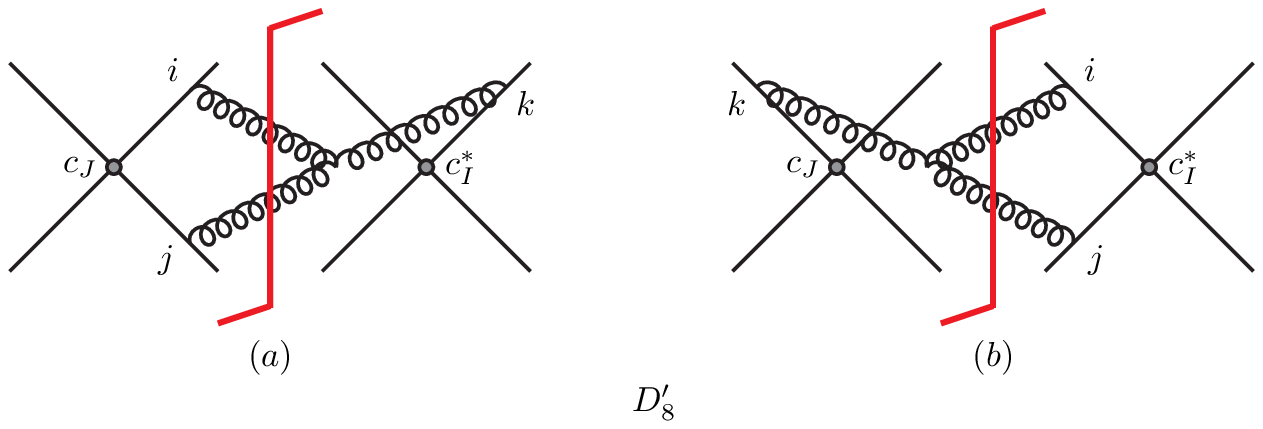}
  \caption{Examples of non-abelian three-Wilson-line integrals
    required in the calculation of the NNLO soft
    matrix. \label{fig:NNLO3Wnonplanar}}
\end{figure}

The three-Wilson-line integrals are of two basic types: the abelian
graphs shown in Figure~\ref{fig:NNLO3Wplanar} and
Figure~\ref{fig:NNLO3W1LV1LR}, and the non-abelian graphs involving a
three-gluon vertex shown in Figure~\ref{fig:NNLO3Wnonplanar}.

We first discuss the abelian graphs. The two diagrams $D_8^{(a)}$ and
$D_8^{(b)}$ in the first row of Figure~\ref{fig:NNLO3Wplanar} are
obviously a convolution product of two NLO functions and introduce no
further computational complications. The sum of these two diagrams
gives a symmetric color structure
\begin{align}
  \label{eq:w8}
  \bm{w}^{(8)}_{ijk} &= \frac{1}{d_R} \, \{ \bm{T}_i^a, \bm{T}_i^b \}
  \, \bm{T}_j^a \, \bm{T}_k^b \, ,
\end{align}
and the integral is
\begin{align}
  \label{eq:I8}
  I_8(\omega,a_{ij},a_{ik}) &= \int [dk] \, [dl] \, \frac{n_i \cdot
    n_j \; n_i \cdot n_k \; \delta(\omega - n_0 \cdot (k+l))}{n_i
    \cdot k \; n_i \cdot l \; n_j \cdot k \; n_k \cdot l} =
  \pi^{2-2\epsilon} \, e^{-2\epsilon\gamma_E} \; \omega^{-1-4\epsilon}
  \; \bar{I}_8(a_{ij},a_{ik}) \, ,
\end{align}
with 
\begin{align}
  \bar{I}_8(a,a') &= \frac{8 e^{2\epsilon\gamma_E} \,
    \Gamma^2(-\epsilon) \, \Gamma(-2\epsilon)}{\Gamma(1-2\epsilon) \,
    \Gamma(1-4\epsilon)} \, (1-a)^{-\epsilon} \, (1-a')^{-\epsilon}
  \nonumber
  \\
  &\hspace{8em} \times {}_2F_1(-\epsilon,-\epsilon,1-\epsilon,a) \,
  {}_2F_1(-\epsilon,-\epsilon,1-\epsilon,a') \, .
\end{align}
Each of the four diagrams in the last two rows of
Figure~\ref{fig:NNLO3Wplanar}, on the other hand, are complicated
functions of two distinct scalar products. However, the sums of the
pairs $(c)+(d)$ and $(e)+(f)$ are proportional to symmetric color
structure $\bm{w}^{(8)}$:
\begin{align}
  D_8^{(c)+(d)} &\to \left( \bm{T}^b_i \, \bm{T}^a_i + \bm{T}^a_i \,
    \bm{T}^b_i \right) \bm{T}^a_j \bm{T}^b_k \int [dk] \, [dl] \,
  \frac{n_i \cdot n_j \; n_i \cdot n_k \; \delta(\omega - n_0 \cdot
    (k+l))}{n_i \cdot l \; n_i \cdot (k+l) \; n_j \cdot k \; n_k \cdot
    l} \, , \nonumber
  \\
  D_8^{(e)+(f)} &\to \left( \bm{T}^a_i \, \bm{T}^b_i + \bm{T}^b_i \,
    \bm{T}^a_i \right) \bm{T}^a_j \bm{T}^b_k \int [dk] \, [dl] \,
  \frac{n_i \cdot n_j \; n_i \cdot n_k \; \delta(\omega - n_0 \cdot
    (k+l))}{n_i \cdot k \; n_i \cdot (k+l) \; n_j \cdot k \; n_k \cdot
    l} \, .
\end{align}
Furthermore, after partial fractioning, the sum of the two integrals
yields the factorized integral (\ref{eq:I8}). Therefore, these abelian
diagrams do not introduce any new calculational complications. In
Appendix~\ref{sec:NAexp} we explain how the non-abelian exponentiation
theorem implies the simple factorized form of the integral multiplying
the symmetric color structure (\ref{eq:w8}). The color matrices for
the $q\bar q$ channel are
\begin{align}
  \bm{w}^{(8)}_{123} &= \bm{w}^{(8)}_{214} = \bm{w}^{(8)}_{314} =
  \bm{w}^{(8)}_{423} = \frac{C_F}{2}
  \begin{pmatrix}
    0 & \frac{N^2-2}{2N}
    \\
    \frac{N^2-2}{2N} & -\frac{N^2-2}{2N^2}
  \end{pmatrix}
   , \nonumber
  \\
  \bm{w}^{(8)}_{124} &= \bm{w}^{(8)}_{213} = \bm{w}^{(8)}_{324} = \bm{w}^{(8)}_{413} = \frac{C_F}{2}
  \begin{pmatrix}
    0 & -\frac{N^2-2}{2N}
    \\
    -\frac{N^2-2}{2N} & -\frac{1}{N^2}
  \end{pmatrix}
  , \nonumber
  \\
  \bm{w}^{(8)}_{134} &= \bm{w}^{(8)}_{234} = \bm{w}^{(8)}_{312} = \bm{w}^{(8)}_{412} = \frac{C_F}{2}
  \begin{pmatrix}
    -2 & -\frac{N^2-4}{2N}
    \\
    -\frac{N^2-4}{2N} & \frac{N^2-3}{2N^2}
  \end{pmatrix}
  ,
\end{align}
while those for the $gg$ channel are 
\begin{align}
  \bm{w}^{(8)}_{123} &= \bm{w}^{(8)}_{214} =\frac{C_A}{2}
  \begin{pmatrix}
    0 & \frac{3}{2} N & 0
    \\
    \frac{3}{2} N & \frac{N^2}{4} & \frac{N^2-4}{4}
    \\
    0 & \frac{N^2-4}{4} & \frac{N^2-4}{4}
    \end{pmatrix}
  , \nonumber
  \\
  \bm{w}^{(8)}_{124} &= \bm{w}^{(8)}_{213} =\frac{C_A}{2}
  \begin{pmatrix}
    0 & -\frac{3}{2} N & 0
    \\
    -\frac{3}{2} N & \frac{N^2}{4} & -\frac{N^2-4}{4}
    \\
    0 & -\frac{N^2-4}{4} & \frac{N^2-4}{4}
  \end{pmatrix}
  , \nonumber
  \\
  \bm{w}^{(8)}_{134} &= \bm{w}^{(8)}_{234} = \bm{w}^{(8)}_{312} =
  \bm{w}^{(8)}_{412} =\frac{C_A}{2}
  \begin{pmatrix}
    -2 & 0 & -\frac{N^2-4}{2 N}
    \\
    0 & -\frac{1}{2} & 0
    \\
    -\frac{N^2-4}{2 N} & 0 & \frac{N^2-4}{2 N^2}
  \end{pmatrix}
  , \nonumber
  \\
  \bm{w}^{(8)}_{314} &= \bm{w}^{(8)}_{423} =\frac{C_A}{2}
  \begin{pmatrix}
    0 & \frac{N^2-2}{2 N} &0
    \\
    \frac{N^2-2}{2 N} & -\frac{1}{4} & -\frac{N^2-4}{4 N^2}
    \\
    0 & -\frac{N^2-4}{4 N^2} & -\frac{N^2-4}{4 N^2}
  \end{pmatrix}
  , \nonumber
  \\
  \bm{w}^{(8)}_{324} &= \bm{w}^{(8)}_{413} =\frac{C_A}{2}
  \begin{pmatrix}
    0 & -\frac{N^2-2}{2 N} &0
    \\
    -\frac{N^2-2}{2 N} & -\frac{1}{4} & \frac{N^2-4}{4 N^2}
    \\
    0 & \frac{N^2-4}{4 N^2} & -\frac{N^2-4}{4 N^2}
  \end{pmatrix} .
\end{align}

We now discuss one-particle cuts of three-Wilson-line abelian
diagrams. These are first of all due to contributions of the type
shown in the first row of Figure~\ref{fig:NNLO3Wplanar}, but where the
connection to parton $k$ is moved to the other side of the cut. These
diagrams factorize in an obvious way into a product of an NLO-type
integral with a scaleless virtual correction and thus evaluate to
zero. The one-particle cuts corresponding to the diagrams in the last
two rows of that figure are more complicated. However, one can show
that the sum of pairs of the type shown in
Figure~\ref{fig:NNLO3W1LV1LR} reduces to the same scaleless integral
mentioned above. Therefore, these one-particle cuts do not contribute
to the soft function.

In addition to the abelian diagrams, we must also evaluate non-abelian
ones of the type shown in Figure~\ref{fig:NNLO3Wnonplanar}. However,
the sum of diagrams in the figure vanishes due to its color structure.
Written explicitly, the two diagrams give
\begin{align}
  D_8^{'(a)} &\to i f^{abc} \, \bm{T}^a_i \, \bm{T}^b_j \,
  \bm{T}^{c}_k \, I'_8 \, ,
  \\
  D_8^{'(b)} &\to -i f^{abc} \, \bm{T}^a_i \, \bm{T}^b_j \,
  \bm{T}^{c}_k \, I'_8 \, ,
\end{align}
where
\begin{align}
  I'_8 &= \int [dk] [dl] \, \frac{n_i^\mu n_j^\nu n_k^\rho \left[
      (2k+l)_\nu g_{\mu \rho} - (k+2l)_\mu g_{\rho \nu} + (l-k)_\rho
      g_{\nu \mu} \right]}{n_i \cdot k \; n_j \cdot l \; n_k \cdot
    (k+l) \; (k+l)^2} \, .
\end{align}
The relative minus sign between the two diagrams originates from the
fact that the gluon propagator carrying momentum $k+l$ appears on
opposite sides of the cut in the two diagrams\footnote{This change of
  sign also occurs in the diagrams generated from the non-abelian two
  Wilson-line diagrams $D_6$ and $D_{7,2}$ in Figure~\ref{fig:NNLO2W}
  by moving the two-gluon attachment to the other side of the cut.
  However, in those cases the order of the two color generators
  $\bm{T}^a_i$ and $\bm{T}^b_i$ is also exchanged, which compensates
  the sign change of the propagators.}. Therefore, the sum of the two
diagrams cancels. This is due to the structure of the color factors
and not the integral itself, and is also true of one-particle cuts not
shown in the figure. Interestingly enough, it would also be true if
the velocity vectors were time-like, even though such anti-symmetric
three-particle correlations are present in the anomalous dimension for
massive particles \cite{Ferroglia:2009ep, Ferroglia:2009ii}. This is
not a contradiction, since the soft function for massive heavy-quark
production obeys an RG equation analogous to (\ref{eq:Sev}) below and
the contribution of the three-particle correlations cancels between
the sum of terms involving $\bm{\gamma}_s$ and
$\bm{\gamma}_s^\dagger$.

\subsection{Four-Wilson-line integrals}
\label{sec:4W}

\begin{figure}
  \begin{center}
    \includegraphics{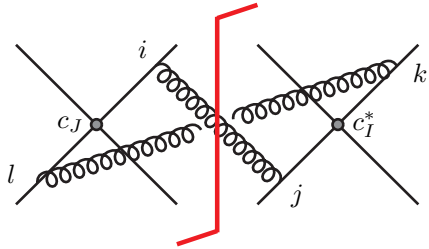}
  \end{center}
  \caption{Four-Wilson-line diagrams contributing to the NNLO soft
    function. \label{fig:NNLO4W}}
\end{figure}

We finally turn to the four-Wilson-line integrals. The two-particle
cuts shown in Figure~\ref{fig:NNLO4W} are a convolution of NLO
integrals and of the same form as $I_4$ and $I_8$. The general
integral reads
\begin{align}
  I_9(\omega,a_{ij},a_{kl}) &= \int [dk] \, [dl] \, \frac{n_i \cdot
    n_j \; n_k \cdot n_l \; \delta(\omega - n_0 \cdot (k+l))}{n_i
    \cdot k \; n_j \cdot k \; n_k \cdot l \; n_l \cdot l} \, ,
\end{align}
and the color factor is
\begin{align}
  \bm{w}^{(9)}_{ij} &= \frac{1}{d_R} \, \bm{T}_i^a \, \bm{T}_j^a \,
  \bm{T}_k^b \, \bm{T}_l^b \; (i \neq j \neq k \neq l) \, .
\end{align}
For two-to-two scattering considered here the relations between the
different scalar products implies that we need only
$I_9(\omega,a_{ij},a_{ij})=I_4(\omega, a_{ij})$, which we already took
into account in writing (\ref{eq:S2bare}). The explicit results for
the color matrices are $\bm{w}^{(9)}_{ij}=\bm{w}^{(3)}_{ij}$ in the
$q\bar q$ channel, and for the $gg$ they read
\begin{align}
  \bm{w}^{(9)}_{12} &= \frac{C_A}{2}
  \begin{pmatrix}
    N^2-1 & 0 & 0
    \\
    0 & -\frac{1}{4} & 0
    \\
    0 & 0 & -\frac{N^2-4}{4N^2}
  \end{pmatrix}
  ,
\end{align}
and $\bm{w}^{(9)}_{ij}=\bm{w}^{(3)}_{ij}$ for $ij=13,\,14$.

Also in this case there are one-particle cuts, but these always
involve a scaleless loop graph and vanish.

\section{Renormalization}
\label{sec:renormalization}

The renormalized soft function is obtained from the bare one by
multiplying on both sides by a matrix-valued UV renormalization
factor. The structure of this renormalization factor follows from the
RG equation for the soft function, which is simplest to discuss at the
level of the Laplace-transformed function. We define this as
\begin{align}
  \label{eq:stilde}
  \tilde{\bm{s}} \left(L, t_1/M^2, \mu \right) &= \int_0^\infty
  d\omega \, \exp{\left(-\frac{\omega}{e^{\gamma_E}\mu
        e^{L/2}}\right)} \, \bm{S}(\omega,t_1/M^2,\mu) \, ,
\end{align}
The integral transform is easily carried out from the bare function
using
\begin{align}
  \int_0^\infty d\omega \exp[-b \omega ]\omega^{-1-n\epsilon}=
  \Gamma(-n\epsilon) b^{\epsilon n} \, .
\end{align}
The RG equation for the Laplace-transformed function was derived in
\cite{Ferroglia:2012ku}, using RG invariance of the cross section and
the results for the hard, fragmentation, and parton luminosities. It
reads
\begin{align}
  \label{eq:Sev}
  \frac{d}{d\ln\mu} \, \tilde{\bm{s}}
  &\left(\ln\frac{M^2}{\mu^2},t_1/M^2,\mu\right) = \nonumber
  \\
  &- \left[ A(\alpha_s) \ln\frac{M^2}{\mu^2} +
    \bm{\gamma}^{s\dagger}(t_1/M^2,\alpha_s) \right] \tilde{\bm{s}}
  \left(\ln\frac{M^2}{\mu^2},t_1/M^2,\mu\right) \nonumber
  \\
  &- \tilde{\bm{s}} \left(\ln\frac{M^2}{\mu^2},t_1/M^2,\mu\right)
  \left[ A(\alpha_s) \ln\frac{M^2}{\mu^2} +
    \bm{\gamma}^s(t_1/M^2,\alpha_s) \right] \, ,
\end{align}
where $A=2C_F\gamma_{\rm cusp}$ in the $q\bar q$ channel and
$A=(C_A+C_F)\gamma_{\rm cusp}$ in the $gg$ channel, and
\begin{align}
  \bm{\gamma}^s(t_1/M^2,\alpha_s) = \bm{\gamma}^h(t_1/M^2,\alpha_s) +
  \left[2\gamma^{\phi}(\alpha_s) + 2\gamma^{\phi_q}(\alpha_s)\right]
  \bm{1} \, .
\end{align}
The non-trivial matrix structure of the soft anomalous dimension is
related to that in the hard function and is expressed through the
function $\bm{\gamma}^h$, which for the different channels is
\begin{align}
  \label{eq:qqmatrix}
  \bm{\gamma}^h_{q\bar{q}} &= 4\gamma^q(\alpha_s) \bm{1} + N
  \gamma_{\text{cusp}}(\alpha_s) \, \left(\ln\frac{-t_1}{M^2}+ i
    \pi\right)
  \begin{pmatrix}
    0 & 0
    \\
    0 & 1
  \end{pmatrix}
  + \gamma_{\text{cusp}}(\alpha_s) \, \ln\frac{t_1^2}{u_1^2}
  \begin{pmatrix}
    0 & \frac{C_F}{2N}
    \\
    1 & -\frac{1}{N}
  \end{pmatrix}
  ,
\end{align}
and
\begin{align}
  \label{eq:ggmatrix}
  \bm{\gamma}^h_{gg} &= \left[ 2\gamma^g(\alpha_s) +
    2\gamma^q(\alpha_s) \right] \bm{1} \nonumber
  \\
  &\quad\mbox{} + N \gamma_{\text{cusp}}(\alpha_s) \, \left(
    \ln\frac{-t_1}{M^2} + i\pi\right)
  \begin{pmatrix}
    0 & 0 & 0
    \\
    0 & 1 & 0
    \\
    0 & 0 & 1
  \end{pmatrix}
  + \gamma_{\text{cusp}}(\alpha_s) \, \ln\frac{t_1^2}{u_1^2}
  \begin{pmatrix}
    0 & \frac{1}{2} & 0
    \\
    1 & -\frac{N}{4} & \frac{N^2-4}{4N}
    \\
    0 & \frac{N}{4} & -\frac{N}{4}
  \end{pmatrix}
  .
\end{align}
Explicit results for the expansion coefficients defined as
\begin{align}
  A(\alpha_s)=\frac{\alpha_s}{4\pi}A_0 +
  \left(\frac{\alpha_s}{4\pi}\right)^2 A_1 + \cdots \, ,
\end{align}
and analogously for the other functions can be found in the the
Appendix of \cite{Ferroglia:2012ku}.

Given the structure of the RG equations, we define the relationship
between the bare and renormalized soft functions according to
\begin{align}
  \label{eq:ren}
  \tilde{\bm{s}}= \bm{Z}_s^{\dagger}\tilde{\bm{s}}_{\rm bare}\bm{Z}_s
  \, .
\end{align}
The bare soft function does not depend on $\mu$. This implies an RG
equation for the renormalization factor which can be integrated to
solve for its explicit expansion in $\epsilon$ (see, e.g.,
\cite{Becher:2009cu}). This yields
\begin{align}
  \label{eq:Zdef}
  \ln{\bm{Z}_s}& =\frac{\alpha_s}{4\pi}\left(-\frac{A_0}{2\epsilon^2}
    +\frac{A_0L+\bm{\gamma}^s_0}{2\epsilon}\right) \nonumber
  \\
  &\hspace{0.5cm} +\left(\frac{\alpha_s}{4\pi}\right)^2 \bigg[\frac{3
    A_0 \beta_0}{8\epsilon^3}+\frac{-A_1-2\beta_0(A_0L+
    \bm{\gamma}^s_0)}{8\epsilon^2}
  +\frac{A_1L+\bm{\gamma}^s_1}{4\epsilon}\ \bigg] +\cdots \, .
\end{align}
Defining expansion coefficients of the renormalization factor and
Laplace-transformed functions in units of $\alpha_s/4\pi$ as in
(\ref{eq:Sexp}), the renormalized NNLO function in Laplace space is
given by
\begin{align}
  \tilde{\bm{s}}^{(2)}(L,t_1/M^2,\mu) &= \tilde{\bm{s}}^{(2)}_{\rm
    bare}+ \bm{Z}_s^{\dagger(2)}\tilde{\bm{s}}^{(0)}+
  \tilde{\bm{s}}^{(0)}\bm{Z}_s^{(2)} \nonumber
  \\
  & \hspace{0.6cm} +\bm{Z}_s^{\dagger(1)}\tilde{\bm{s}}^{(1)}_{\rm
    bare} + \tilde{\bm{s}}^{(1)}_{\rm bare}\bm{Z}_s^{(1)}+
  \bm{Z}_s^{\dagger(1)} \tilde{\bm{s}}^{(0)}\bm{Z}_s^{(1)}
  -\frac{\beta_0}{\epsilon}\tilde{\bm{s}}^{(1)}_{\rm bare} \,.
\end{align}
Evaluating this equation, we find that the renormalized function on
the left-hand side is indeed finite in the limit $\epsilon\to 0$. This
shows that the renormalization factor (\ref{eq:Zdef}) following from
the RG equation deduced from the factorization formula
(\ref{eq:cartoonfact}) is correct, or can otherwise be viewed as a
cross-check on our calculations. The explicit expressions for the
renormalized functions are rather lengthy and we relegate them to
Appendix~\ref{sec:Results}. The terms proportional to powers of $L$ in
the renormalized function are in agreement with the approximate NNLO
formulas mentioned in \cite{Ferroglia:2012ku}, while the
$L$-independent pieces are new.

\newpage

\section{Conclusions}
\label{sec:conclusions}

We evaluated the NNLO corrections to the soft function needed to
describe the pair invariant mass distribution in top-quark pair
production at hadron colliders in the small-mass limit. At the
technical level, this required us to obtain real-emission type
corrections related to a product of Wilson-loop operators depending on
four light-like Wilson lines. This is the first NNLO calculation of a
soft function which involves non-trivial matrix structure in color
space. We showed that the IR structure of the bare function is
consistent with known expressions for the two-loop anomalous dimension
matrix derived in \cite{Ferroglia:2012ku}. The final results, given in
the Appendix, turned out to be rather simple. This is because in the
sum of all diagrams contributions from integrals involving three or
more Wilson lines multiply a color structure whose coefficient is
constrained by the non-abelian exponentiation theorem to be a product
of NLO integrals. The non-abelian three-parton graphs are not
constrained by non-abelian exponentiation, but these vanish after
summing over all diagrams.

Combined with known NNLO results for heavy-quark fragmentation
functions and virtual corrections to two-to-two processes, our
calculations will allow an evaluation of the invariant mass
distribution in the small-mass limit at the level of a full virtual
plus soft approximation. This will provide valuable information
concerning the importance of higher-order corrections to NLO+NNLL
predictions from \cite{Ahrens:2010zv}. Moreover, by an appropriate
modification of the delta-function constraint in the definition
(\ref{eq:Sdef}), we can immediately obtain the soft function needed to
study single-particle distributions in the $p_T$ and rapidity of the
top quark to the same level of accuracy, opening up the possibility of
studying higher-order corrections to the results of
\cite{Kidonakis:2010dk,Ahrens:2011mw}. Finally, we anticipate
applications for threshold resummation in dijet production.

\section*{Acknowledgments}
We would like to thank Matthias Neubert for useful discussions. This
research was supported in part by the PSC-CUNY Award N. 64133-00 42,
by the NSF grant PHY-1068317, by the German Research Foundation under
grant NE398/3-1, 577401: {\em Applications of Effective Field Theories
  to Collider Physics}, by the European Commission through the {\em
  LHCPhenoNet\/} Initial Training Network PITN-GA-2010-264564, and by
the Schweizer Nationalfonds under grant 200020-141360/1.

\appendix

\section{Calculation of $I_5$ as an example}
\label{sec:Example}

Written out explicitly, the integral $I_5$ reads (with $\delta^+(k^2)=
\delta(k^2)\theta(k^0)$)
\begin{align}
  I_5(\omega,a_{ij}) &= \int d^dl_1 \, d^dl_2 \; \frac{(n_i \cdot
    n_j)^2 \; \delta^+(l_1^2) \; \delta^+(l_2^2) \; \delta(\omega -
    n_0 \cdot (l_1+l_2))}{n_i \cdot l_1 \; n_i \cdot (l_1+l_2) \; n_j
    \cdot l_1 \; n_j \cdot (l_1+l_2)} \nonumber
  \\
  &= \int d^dk \; \frac{n_i \cdot n_j \; \delta(\omega - n_0 \cdot
    k)}{n_i \cdot k \; n_j \cdot k} \int d^dl_1 \, d^dl_2 \; \frac{n_i
    \cdot n_j \; \delta^+(l_1^2) \; \delta^+(l_2^2) \;
    \delta^{(d)}(k-l_1-l_2)}{n_i \cdot l_1 \; n_j \cdot l_1} \nonumber
  \\
  &= \int d^dk \; \frac{n_i \cdot n_j \; \delta(\omega - n_0 \cdot
    k)}{n_i \cdot k \; n_j \cdot k} \; 2\pi^{1-\epsilon} \;
  \frac{\Gamma(-\epsilon)}{\Gamma(1-2\epsilon)} \left( \frac{2 \; n_i
      \cdot k \; n_j \cdot k}{n_i \cdot n_j} \right)^\epsilon
  \nonumber
  \\
  &\hspace{5em} \times \left( k^2 \right)^{-1-2\epsilon} \; {}_2F_1
  \bigg( -\epsilon, -\epsilon, 1-\epsilon, \frac{n_i \cdot n_j \;
    k_T^2}{2 \; n_i \cdot k \; n_j \cdot k} \bigg) \, .
\end{align}
To derive the equality in the third line we used the auxiliary
integral given in \cite{Li:2011zp, Becher:2012za}. The remaining
$d$-dimensional integral over $k$ can be solved using the steps
outlined in \cite{Becher:2012za}. We first define light-cone
coordinates
\begin{gather*}
  k^\mu = k_+ \, \frac{n_i^\mu}{\sqrt{2 \, n_i \cdot n_j}} + k_- \,
  \frac{n_j^\mu}{\sqrt{2 \, n_i \cdot n_j}} + k_\perp^\mu \, ,
  \\
  k_+ = \frac{n_j \cdot k}{\sqrt{n_i \cdot n_j / 2}} \, , \quad k_- =
  \frac{n_i \cdot k}{\sqrt{n_i \cdot n_j / 2}} \, , \quad k_T^2 =
  -k_\perp^2 \, ,
\end{gather*}
and use these to parameterize the integral over $k$. This yields
\begin{align}
  I_5(\omega,a_{ij}) &= \frac{2^{2-2\epsilon} \, \pi^{1-2\epsilon} \,
    \Gamma(-\epsilon) \, \Gamma(1-\epsilon)}{\Gamma^2(1-2\epsilon)}
  \int dk_+dk_- \int_0^\infty dk_T \, k_T^{1-2\epsilon} \int_{-1}^1
  d\cos\theta \, \sin^{-1-2\epsilon}\theta \nonumber
  \\
  &\hspace{2em} \times (k_+k_-)^{-1+\epsilon} \,
  (k_+k_--k_T^2)^{-1-2\epsilon} \, {}_2F_1 \bigg( -\epsilon,
  -\epsilon, 1-\epsilon, \frac{k_T^2}{k_+k_-} \bigg) \nonumber
  \\
  &\hspace{2em} \times \delta \bigg( \omega -
  \frac{k_+n_{0-}+k_-n_{0+}}{2} + n_{0T}k_T\cos\theta \bigg) \, .
\end{align}
We then make the change of variables
\begin{gather*}
  k_+ = \frac{\omega}{n_{0-}} \, x \, y \, , \quad k_- =
  \frac{\omega}{n_{0+}} \, (1-x) \, y \, , \quad dk_+dk_- =
  \frac{\omega^2}{n_{0+}n_{0-}} \, y \, dxdy \, ,
  \\
  k_T = \frac{\omega}{\sqrt{n_{0+}n_{0-}}} \, \sqrt{x(1-x)} \, y \, u
  \, , \quad dk_T = \frac{\omega}{\sqrt{n_{0+}n_{0-}}} \,
  \sqrt{x(1-x)} \, y \, du \, ,
\end{gather*}
and express the light-cone components of $n_0$ in terms of $a_{ij}$ to
arrive at
\begin{align}
  I_5(\omega,a_{ij}) &= \frac{2^{2+2 \epsilon} \, \pi^{1-2\epsilon} \,
    \Gamma(-\epsilon) \, \Gamma(1-\epsilon)}{\Gamma^2(1-2\epsilon)} \;
  \omega^{-1-4\epsilon} \, (1-a_{ij})^{-2\epsilon} \nonumber
  \\
  &\hspace{-2em} \times \int_0^1 dx \int_0^\infty dy \int_{-1}^1
  d\cos\theta \, \sin^{-1-2\epsilon}\theta \int_0^1 du \,
  [x(1-x)]^{-1-2\epsilon} \, y^{-1-4\epsilon} \, u^{1-2\epsilon} \,
  (1-u^2)^{-1-2\epsilon} \nonumber
  \\
  &\hspace{-2em} \times {}_2F_1(-\epsilon,-\epsilon,1-\epsilon,u^2) \;
  \delta \bigg( 1 - \frac{y}{2} + \sqrt{a_{ij}x(1-x)} \, y \, u \,
  \cos\theta \bigg) \, .
\end{align}
We next perform the $y$-integration using the delta-function
constraint and integrate over $\cos\theta$, generating a
hypergeometric function written in the first equality below. After
that we perform a change of variables to the argument of this
hypergeometric function, integrate, and perform another change of
variables to the argument of the original hypergeometric function.
Explicitly,
\begin{align}
  I_5(\omega,a_{ij}) &= \frac{4 \, \pi^{2-2\epsilon} \,
    \Gamma(-\epsilon)}{\Gamma(1-\epsilon) \, \Gamma(1-2\epsilon)} \;
  \omega^{-1-4\epsilon} \, (1-a)^{-2\epsilon} \nonumber
  \\
  &\times \int_0^1 du \, u^{1-2\epsilon} \, (1-u^2)^{-1-2\epsilon} \;
  {}_2F_1(-\epsilon,-\epsilon,1-\epsilon,u^2) \nonumber
  \\
  &\times \int_0^1 dx \, [x(1-x)]^{-1-2\epsilon} \; {}_2F_1 \bigg(
  \frac{1}{2}-2\epsilon, -2\epsilon, 1-\epsilon, 4a_{ij}x(1-x)u^2
  \bigg) \nonumber
  \\
  &= \frac{2^{3+4\epsilon} \, \pi^{2-2\epsilon} \,
    \Gamma(-\epsilon)}{\Gamma(1-\epsilon) \, \Gamma(1-2\epsilon)} \;
  \omega^{-1-4\epsilon} \, (1-a_{ij})^{-2\epsilon} \nonumber
  \\
  &\times \int_0^1 du \, u^{1-2\epsilon} \, (1-u^2)^{-1-2\epsilon} \;
  {}_2F_1(-\epsilon,-\epsilon,1-\epsilon,u^2) \nonumber
  \\
  &\times \int_0^1 dv \, v^{-1-2\epsilon} \, (1-v)^{-1/2} \; {}_2F_1
  \bigg( \frac{1}{2}-2\epsilon, -2\epsilon, 1-\epsilon, a_{ij}u^2v
  \bigg) \nonumber
  \\
  &= \frac{4 \, \pi^{2-2\epsilon} \, \Gamma(-\epsilon) \,
    \Gamma(-2\epsilon)}{\Gamma(1-\epsilon) \, \Gamma(1-4\epsilon)} \;
  \omega^{-1-4\epsilon} \, (1-a_{ij})^{-2\epsilon} \nonumber
  \\
  &\times \int_0^1 dt \, t^{-\epsilon} \, (1-t)^{-1-2\epsilon} \;
  {}_2F_1(-\epsilon,-\epsilon,1-\epsilon,t) \;
  {}_2F_1(-2\epsilon,-2\epsilon,1-\epsilon,a_{ij}t) \, .
\end{align}
The integral over the hypergeometric functions is sufficiently
complicated that we evaluate it as an expansion in $\epsilon$. This is
achieved using the standard identity
\begin{equation}
  (1-t)^{-1 - n \epsilon}= -\frac{1}{n\epsilon}\delta(1-t)
  +\sum_{m=0}^{\infty}\frac{(-n \epsilon)^m}{m!}
  \left[\frac{\ln^m(1-t)}{1-t}\right]_+ \, ,
\end{equation}
and evaluating the integrals over the plus distributions after
expanding the other parts of the integral in $\epsilon$. This leads to
parametric integrals over HPLs. Some of these are straightforward, and
even the most difficult ones are not hard to handle by deriving and
solving differential equations with respect to the parameter $a_{ij}$,
yielding the result given in Section~\ref{sec:NNLObare}.

\section{Constraints from  non-abelian exponentiation}
\label{sec:NAexp}

In this section we briefly explain the consistency of our results with
the non-abelian exponentiation theorem \cite{Gatheral:1983cz,
  Frenkel:1984pz}. In general, this theorem states that the soft
function can be written as the exponential of a simpler quantity. In
particular, in the language of \cite{Frenkel:1984pz}, the exponent
receives contributions only from diagrams involving single connected
webs. For us, the important point is that this theorem implies that
after summing over all diagrams the coefficients of certain color
factors in the NNLO function are determined by exponentiating the NLO
result. This exponentiation occurs in position or Laplace space rather
than momentum space. We find it convenient to work with the
Laplace-space function (\ref{eq:stilde}). Our procedure is to take the
exponential of the one-loop bare function and expand it to second
order:
\begin{align}
  \label{eq:expansion}
  \frac{1}{d_R} \exp \left[ - \sum_{i,j} \bm{T}_i \cdot \bm{T}_j \,
    \tilde{I}_{1}(L,a_{ij}) \right] &= \frac{1}{d_R} \Bigg[ 1 -
  \sum_{i,j} \bm{T}_i \cdot \bm{T}_j \, \tilde{I}_1(L,a_{ij})
  \nonumber \\
  &\hspace{0.3cm} + \frac{1}{2} \sum_{i,j,k,l} \bm{T}_i \cdot \bm T_j
  \; \bm{T}_k \cdot \bm{T}_l \; \tilde{I}_1(L,a_{ij}) \,
  \tilde{I}_1(L,a_{kl}) + \cdots \Bigg] \, .
\end{align}
The term on the second line produces two, three and four parton terms
whose structure must be reproduced in our explicit results. First, we
can write the two parton terms as
\begin{align}
  \frac{2}{ d_R} \Bigl( \bm{T}^a_1\cdot \bm{T}^a_2 \bm{T}^b_1 \cdot
  \bm{T}^b_2 \tilde{I}^2_{1}(L,a_{12}) + \bm{T}^a_3\cdot \bm{T}^a_4
  \bm{T}^b_3 \cdot \bm{T}^b_4
  \tilde{I}^2_{1}(L,a_{12})\nonumber \\
  + 2 \bm{T}^a_1\cdot \bm{T}^a_3 \bm{T}^b_1 \cdot \bm{T}^b_3
  \tilde{I}^2_{1}(L,a_{13}) + 2 \bm{T}^a_1\cdot \bm{T}^a_4 \bm{T}^b_1
  \cdot \bm{T}^b_4 \tilde{I}^2_{1}(L,a_{14})\Bigr) \, .
\end{align}
We thus expect the coefficient of $\bm{w}^{(3)}_{ij}$ to be restricted
by the non-abelian exponentiation theorem, and a relation between
$\tilde{I}_3$, $\tilde{I}_4$, $\tilde{I}_5$ and the NLO integral.
Indeed, we have already made explicit in (\ref{eq:Iset}) that
$I_3=I_4/2-I_5$, and one can easily verify that $\tilde{I}_4
=\tilde{I}_1^2$. After rewriting the color factor $\bm{w}^{(4)}_{ij}$
as in the last equality in (\ref{eq:2WT}) and performing the sum over
legs one can then show that the coefficient of $\bm{w}^{(3)}_{ij}$ in
(\ref{eq:Legs}) is proportional to the factorized integral
$\tilde{I}_4$, as required. The remaining two-Wilson-line integrals
are proportional to the NLO matrix $\bm{w}_{ij}^{(1)}$ and are thus
single connected webs which contribute to the second-order expansion
of the exponent directly.

Next we deal with three-parton terms. To understand their structure we
consider the concrete example where two gluons attach to the parton
with velocity $n_1$ and the other two to the partons with velocity
$n_2$ and $n_3$. The contribution of these terms to
(\ref{eq:expansion}) is
\begin{align}  
  \frac{2}{d_R} \left( \bm{T}^a_1\cdot \bm{T}^a_2 \bm{T}^b_1 \cdot
    \bm{T}^b_3 + \bm{T}^a_1\cdot \bm{T}^a_3 \bm{T}^b_1 \cdot
    \bm{T}^b_2\right) \tilde{I}_{1}(L,a_{12})\tilde{I}_{1}(L,a_{13}) =
  2\bm{w}_{123}^{(8)}\tilde{I}_{1}(L,a_{12})\tilde{I}_{1}(L,a_{13}) ,
\end{align}
where we have used that the color matrices for different partons
commute. The above relation implies that
$\tilde{I}_1(L,a_{12})\tilde{I}_1(L,a_{13}) =
\tilde{I}_8(L,a_{12},a_{13})$, and one can confirm that this is indeed
the case. Including all permutations in the expansion of the
exponential, we reproduce the contributions proportional to the
$\bm{w}_{ijk}^{(8)}$ in (\ref{eq:Legs}).

Finally, we turn to the four-parton terms, considering as an example
the case where the two partons with velocity $n_1$ and $n_2$ are
connected to each other. These contribute to (\ref{eq:expansion}) as
\begin{align}
  \frac{4}{d_R} \bm{T}^a_1\cdot \bm{T}^a_2 \bm{T}^b_3 \cdot \bm{T}^b_4
  \tilde{I}_1(a_{12})\tilde{I}_1(a_{34}) =
  4\bm{w}^{(9)}_{12}\tilde{I}_4(a_{12}) \,.
\end{align}
This accounts explicitly for the four-parton term
$\bm{w}^{(9)}_{12}I_4(a_{12})$ (\ref{eq:Legs}). The remaining
four-parton terms appear as $ \bm{w}^{(3)}_{ij}I_4(a_{ij})$.

\section{Renormalized Soft Functions}
\label{sec:Results}

We list here the results for the renormalized soft function in the
$q\bar q$ and $gg$ production channels. For the sake of brevity we set
$N=3$ and take into account that the soft function is symmetric. In
the following, we indicate the element $i j$ of the matrix
$\tilde{\bm{s}}^{(n)}_{k}$ ($k \in \{q\bar{q}, gg\}$) as
$\tilde{s}^{(n)}_{k \, ij}$.

\subsection{Quark Annihilation Channel}

The elements of the NLO soft matrix in Laplace space are

\begin{align}
  \tilde{s}^{(1)}_{qq\, 11} &= 16 L^2 + \frac{8}{3} \pi^2 \, , \nonumber \\
  \tilde{s}^{(1)}_{qq\, 12} &= \frac{8 \pi ^2}{9}+ \frac{16}{3} L
  H_{0}(x_t)+\frac{16}{3} L H_{1}(x_t)-\frac{16}{3}
  H_{2}(x_t)+\frac{16}{3} H_{0,0}(x_t)+\frac{16}{3}
  H_{1,0}(x_t)-\frac{16}{3} H_{1,1}(x_t)\, , \nonumber \\
  \tilde{s}^{(1)}_{qq\, 22} &=\frac{32 L^2}{9}+\frac{44 \pi
    ^2}{27}+\frac{56}{9} L H_{0}(x_t)-\frac{16}{9} L
  H_{1}(x_t)+\frac{16}{9} H_{2}(x_t)+\frac{56}{9} H_{0,0}(x_t)
  +\frac{56}{9} H_{1,0}(x_t)\nonumber \\ & +\frac{16}{9} H_{1,1}(x_t)
  \, .
\end{align}
The elements of the NNLO soft matrix in Laplace space are 
\begin{align}
\tilde{s}^{(2)}_{qq\, 11} &= \frac{19424}{27} - \frac{6464}{9} L + \frac{1072}{3} L^2 - \frac{176}{3} L^3 + 
\frac{128}{3} L^4 - 
\frac{2624}{81} N_l + \frac{896}{27} L N_l - \frac{160}{9} L^2 N_l \nonumber \\ &
+ \frac{32}{9} L^3 N_l + 
       \frac{268}{9}   \pi^2 - \frac{16}{9} L^2   \pi^2 - \frac{40}{27} N_l   \pi^2 - \frac{56}{9}  \pi^4  +
       \frac{64}{9} L   \pi^2 H_{0}(x_t) + \frac{64}{9} L  \pi^2 H_{1}(x_t) 
       \nonumber \\ &+ 
       \Biggl[\frac{128}{3} L^2 - \frac{64}{9}  \pi^2 \Biggr] H_{2}(x_t) - 
       \frac{128}{3} L H_{3}(x_t) - 128 H_{4}( x_t) + \Biggl[\frac{128}{3} L^2 
+ \frac{64}{9}   \pi^2 \Biggr] H_{0, 0}(x_t)  \nonumber \\ &+ \frac{128}{3} L^2 H_{1, 0}(x_t) + 
\frac{64}{9}   \pi^2 H_{1, 0}(x_t) + \Biggl[\frac{128}{3} L^2  - 
       \frac{64}{9}   \pi^2 \Biggr] H_{1, 1}(x_t) - \frac{128}{3} L H_{1, 2}( x_t)  \nonumber \\ &- 
       128 H_{1, 3}( x_t) + \frac{128}{3} L H_{2, 0}(x_t) - 
       128 L H_{2, 1}(x_t) - \frac{128}{3} H_{2, 2}(x_t) - 
       \frac{128}{3} H_{3, 0}(x_t) \nonumber \\ &+ \frac{128}{3} H_{3, 1}(x_t)  + 
       128 L H_{0, 0, 0}(x_t) + 128 L H_{1, 0, 0}(x_t) + 
       \frac{128}{3} L H_{1, 1, 0}(x_t) \nonumber \\ &- 128 L H_{1, 1, 1}(x_t) - 
       \frac{128}{3} H_{1, 1, 2}(x_t) - \frac{128}{3} H_{1, 2, 0}(x_t)  + 
       \frac{128}{3} H_{1, 2, 1}(x_t) + \frac{128}{3} H_{2, 0, 0}(x_t)\nonumber \\ &  - 
       128 H_{2, 1, 0}(x_t) + 128 H_{2, 1, 1}(x_t) + 
       128 H_{0, 0, 0, 0}( x_t) + 128 H_{1, 0, 0, 0}(x_t)  + 
       \frac{128}{3} H_{1, 1, 0, 0}(x_t) \nonumber \\ & - 128 H_{1, 1, 1, 0}(x_t)  + 
       128 H_{1, 1, 1, 1}(x_t) - \frac{176}{3} \zeta_3 + 672 L \zeta_3 + 
       \frac{32}{9} N_l \zeta_3 \, ,\nonumber \\
\tilde{s}^{(2)}_{qq\, 12} &=  \frac{536}{27}  \pi^2 - \frac{88}{9} L   \pi^2 + \frac{128}{27} L^2   \pi^2  - 
       \frac{80}{81} N_l   \pi^2 + \frac{16}{27} L N_l   \pi^2 - \frac{124}{135}  \pi^4 + \Biggl[ - 
       \frac{3232}{27} 
       \nonumber \\ &+ \frac{1072}{9} L  - 
       \frac{88}{3} L^2  + \frac{256}{9} L^3   + 
       \frac{448}{81} N_l  - \frac{160}{27} L N_l  
       + 
       \frac{16}{9} L^2 N_l  - \frac{88}{9}   \pi^2  
       + 
       \frac{208}{27} L   \pi^2  \nonumber \\ &+ \frac{16}{27} N_l   \pi^2 + 112  \zeta_3 \Biggr] H_{0}( x_t)  + \Biggl[- 
       \frac{3232}{27}  
       + \frac{1072}{9} L  - 
       \frac{88}{3} L^2  + \frac{256}{9} L^3  + 
       \frac{448}{81} N_l  - \frac{160}{27} L N_l   \nonumber \\ &+ 
       \frac{16}{9} L^2 N_l  - \frac{88}{9}   \pi^2  + 
       \frac{64}{27} L   \pi^2  + \frac{16}{27} N_l   \pi^2 + 
              144  \zeta_3 \Biggr] H_{1}(x_t) 
        +\Biggl[- \frac{1072}{9} 
       + \frac{176}{3} L  
       \nonumber \\ &- 
       \frac{32}{3} L^2  + \frac{160}{27} N_l  - 
       \frac{32}{9} L N_l  - \frac{64}{27}   \pi^2 \Biggr]H_{2}(x_t) 
       +\Biggl[ - 
       \frac{176}{3}  - \frac{160}{9} L   + 
       \frac{32}{9} N_l \Biggr] H_{3}(x_t) 
       \nonumber \\ & - \frac{64}{3} H_{4}( x_t) + \Biggl[\frac{1072}{9} 
        - \frac{176}{3} L  + \frac{704}{9} L^2  - 
       \frac{160}{27} N_l + \frac{32}{9} L N_l 
       + 
       \frac{208}{27}   \pi^2 \Biggr]H_{0, 0}(x_t)
        \nonumber \\ &+ \Biggl[\frac{1072}{9}  - 
       \frac{176}{3} L  + \frac{416}{9} L^2  
       - 
       \frac{160}{27} N_l  
       + \frac{32}{9} L N_l 
        + \frac{64}{27}   \pi^2\Biggr] H_{1, 0}(x_t)  + \Biggl[- \frac{1072}{9}  \nonumber \\ & + 
       \frac{176}{3} L 
       - \frac{128}{3} L^2  
      + \frac{160}{27} N_l  - \frac{32}{9} L N_l 
      - 
       \frac{208}{27}   \pi^2 \Biggr] H_{1, 1}(x_t) 
        + \Biggl[- \frac{176}{3} + 
       \frac{128}{9} L  
       \nonumber \\ & + \frac{32}{9} N_l \Biggr] H_{1, 2}(x_t) - 
       \frac{160}{3} H_{1, 3}(x_t)  + \Biggl[- \frac{176}{3}   + 
       \frac{448}{9} L + \frac{32}{9} N_l \Biggr] H_{2, 0}(x_t) 
       \nonumber \\ & + \Biggl[- 
       \frac{176}{3} + \frac{32}{3} L  +
       \frac{32}{9} N_l \Biggr] H_{2, 1}(x_t) - \frac{160}{9} H_{2, 2}(x_t)  + 
       \frac{128}{9} H_{3, 0}(x_t) + \frac{160}{9} H_{3, 1}(x_t)
       \nonumber \\ &  + \Biggl[- 
       \frac{176}{3}  + \frac{448}{3} L + 
       \frac{32}{9} N_l\Biggr] H_{0, 0, 0}(x_t)+ \Biggl[ - \frac{176}{3}   + 
       \frac{352}{3} L 
       + \frac{32}{9} N_l \Biggr] H_{1, 0, 0}(x_t) \nonumber \\ &
       +\Biggl[ - \frac{176}{3} + \frac{160}{9} L + 
       \frac{32}{9} N_l  \Biggr] H_{1, 1, 0}(x_t) 
       +\Biggr[- \frac{176}{3}   + 
       \frac{128}{3} L  + \frac{32}{9} N_l \Biggl] H_{1, 1, 1}(x_t) \nonumber \\ &- 
       \frac{448}{9} H_{1, 1, 2}(x_t) 
       - \frac{160}{9} H_{1, 2, 0}(x_t) - 
       \frac{416}{9} H_{1, 2, 1}(x_t) + \frac{736}{9} H_{2, 0, 0}(x_t)  - 
       \frac{160}{3} H_{2, 1, 0}(x_t)
       \nonumber \\ &+ \frac{64}{3} H_{2, 1, 1}(x_t) 
       + 
       \frac{448}{3} H_{0, 0, 0, 0}(x_t) + \frac{256}{3} H_{1, 0, 0, 0}(x_t) + 
       \frac{160}{9} H_{1, 1, 0, 0}(x_t) \nonumber \\ &- \frac{256}{3} H_{1, 1, 1, 0}(x_t)  - 
       \frac{128}{3} H_{1, 1, 1, 1}(x_t)   \, ,\nonumber \\
\tilde{s}^{(2)}_{qq\, 22} &= \frac{38848}{243} - \frac{12928}{81}L + \frac{2144}{27} L^2- 
       \frac{352}{27} L^3 + \frac{256}{27} L^4 - \frac{5248}{729} N_l + \frac{1792}{243} L N_l \nonumber \\ &- 
       \frac{320}{81} L^2 N_l + \frac{64}{81} L^3 N_l + \frac{268}{9} \pi^2 - \frac{308}{27} L \pi^2 + 
       \frac{416}{81} L^2 \pi^2 - \frac{40}{27} N_l \pi^2 + \frac{56}{81} L N_l \pi^2 
       \nonumber \\ & - 
       \frac{994}{405} \pi^4 + \Biggl[- \frac{11312}{81} + \frac{3752}{27} L  - 
       \frac{308}{9} L^2  + \frac{896}{27} L^3  + 
       \frac{1568}{243} N_l  - \frac{560}{81} L N_l + 
       \frac{56}{27} L^2 N_l  \nonumber \\ &- \frac{308}{27} \pi^2  + 
       \frac{856}{81} L \pi^2  + \frac{56}{81} N_l \pi^2  + 
              \frac{392}{3}   \zeta_3\Biggr]H_{0}(x_t) + 
       \Biggl[\frac{3232}{81}  - \frac{1072}{27} L  + 
       \frac{88}{9} L^2 - \frac{256}{27} L^3  \nonumber \\ &- 
       \frac{448}{243} N_l  + \frac{160}{81} L N_l  - 
       \frac{16}{27} L^2 N_l  - \frac{308}{27} \pi^2  - 
       \frac{80}{81} L \pi^2  + \frac{56}{81} N_l \pi^2 \Biggr] H_{1}(x_t) + 
       \Biggl[\frac{1072}{27}  
       \nonumber  \\ & 
       - \frac{176}{9} L + 
       \frac{64}{27} L^2  - \frac{160}{81} N_l  + 
       \frac{32}{27} L N_l  +\frac{80}{81} \pi^2 \Biggr] H_{2}(x_t) + 
       \Biggl[\frac{176}{9}  + \frac{64}{9} L  - 
       \frac{32}{27} N_l \Biggr] H_{3}(x_t) 
       \nonumber \\ &
       + \frac{32}{3} H_{4}(x_t)+ \Biggl[
       \frac{3752}{27} - \frac{616}{9} L  + 
       \frac{2720}{27} L^2 - \frac{560}{81} N_l + 
       \frac{112}{27} L N_l  + \frac{856}{81} \pi^2 \Biggr]H_{0, 0}(x_t) 
       \nonumber \\ &+  \Biggl[
       \frac{3752}{27}  - \frac{616}{9} L  + 
      \frac{704}{27} L^2 - \frac{560}{81} N_l  + 
       \frac{112}{27} L N_l  + \frac{352}{81} \pi^2\Biggr] H_{1, 0}(x_t) + \Biggl[
       \frac{1072}{27} - \frac{176}{9} L  \nonumber \\ &+ 
       \frac{640}{27} L^2  - \frac{160}{81} N_l  + 
       \frac{32}{27} L N_l  - \frac{424}{81} \pi^2  \Biggr]H_{1, 1}(x_t) + \Biggl[
       \frac{176}{9}  - \frac{128}{9} L  - 
       \frac{32}{27} N_l \Biggr] H_{1, 2}(x_t) 
       \nonumber \\ &+ \frac{64}{3} H_{1, 3}(x_t) +  
       \Biggl[-\frac{616}{9} + \frac{608}{9} L  + 
       \frac{112}{27} N_l \Biggr] H_{2, 0}(x_t) + \Biggl[\frac{176}{9}  - 
       \frac{64}{3} L  - \frac{32}{27} N_l \Biggr]H_{2, 1}(x_t) 
       \nonumber \\ & + 
       \frac{160}{9} H_{2, 2}(x_t)+ \frac{736}{9} H_{3, 0}(x_t) + 
       \frac{128}{9} H_{3, 1}(x_t) +\Biggl[- \frac{616}{9} + 
       \frac{608}{3} L + \frac{112}{27} N_l \Biggr] H_{0, 0, 0}(x_t) \nonumber \\ & 
       +\Biggl[- \frac{616}{9} + 128 L  + \frac{112}{27} N_l \Biggr]H_{1, 0, 0}(x_t) + \Biggl[- \frac{616}{9}  - 
       \frac{64}{9} L  + \frac{112}{27} N_l \Biggr] H_{1, 1, 0}(x_t) 
       \nonumber \\ &
       + 
       \Biggl[ \frac{176}{9} - \frac{128}{3} L  - 
       \frac{32}{27} N_l \Biggr] H_{1, 1, 1}(x_t) + \frac{256}{9} H_{1, 1, 2}(x_t) + 
       \frac{400}{9} H_{1, 2, 0}(x_t) + \frac{320}{9} H_{1, 2, 1}(x_t) \nonumber \\ &+ 
       \frac{1280}{9} H_{2, 0, 0}(x_t) + \frac{64}{3} H_{2, 1, 0}(x_t) + 
       \frac{64}{3} H_{2, 1, 1}(x_t) 
       + \frac{608}{3} H_{0, 0, 0, 0}(x_t) + 
       128 H_{1, 0, 0, 0}(x_t) \nonumber \\ &+ \frac{608}{9} H_{1, 1, 0, 0}(x_t) - 
       16 H_{1, 1, 1, 0}(x_t) + \frac{128}{3} H_{1, 1, 1, 1}(x_t) - 
       \frac{352}{27}  \zeta_3 + \frac{448}{3} L  \zeta_3 + \frac{64}{81} N_l  \zeta_3  \, .
\end{align}

\subsection{Gluon Fusion Channel}

The elements of the NLO soft matrix in Laplace space are

\begin{align}
\tilde{s}^{(1)}_{gg\, 11} &= 26 L^2+\frac{13}{3}\pi ^2\, ,\nonumber \\
\tilde{s}^{(1)}_{gg\, 12} &= 2 \pi ^2+12 L H_{0}(x_t)+12 L H_{1}(x_t)-12 H_{2}(x_t)+12
   H_{0,0}(x_t)+12 H_{1,0}(x_t)-12 H_{1,1}(x_t) \, ,\nonumber \\
\tilde{s}^{(1)}_{gg\, 13} &= 0 \, ,\nonumber \\
\tilde{s}^{(1)}_{gg\, 22} &= 13 L^2+\frac{11}{3} \pi ^2+9 L H_{0}(x_t)-9 L H_{1}(x_t)+9 H_{2}(x_t)+9 H_{0,0}(x_t)+9
   H_{1,0}(x_t)+9 H_{1,1}(x_t)\, ,\nonumber \\
\tilde{s}^{(1)}_{gg\, 23} &= \frac{5}{6} \pi ^2+5 L H_{0}(x_t)+5 L H_{1}(x_t)-5 H_{2}(x_t)+5 H_{0,0}(x_t)+5
   H_{1,0}(x_t)-5 H_{1,1}(x_t)\, ,\nonumber \\
\tilde{s}^{(1)}_{gg\, 33} &= \frac{55}{27}  \pi ^2+\frac{65 L^2}{9}+5 L H_{0}(x_t)-5 L H_{1}(x_t)+5 H_{2}(x_t)+5 H_{0,0}(x_t)+5
   H_{1,0}(x_t)+5 H_{1,1}(x_t)\, . 
\end{align}
The elements of the NNLO soft matrix in Laplace space are
\begin{align}
\tilde{s}^{(2)}_{gg\, 11} &=\frac{31564}{27} -\frac{10504
   }{9} L+\frac{1742 }{3} L^2-\frac{286}{3} L^3+\frac{338}{3} L^4-\frac{4264 }{81} N_l
   +\frac{1456 }{27}L N_l-\frac{260}{9} L^2 N_l
\nonumber \\ & +\frac{52}{9} L^3 N_l+\frac{871}{18}  \pi ^2+\frac{104}{9}  \pi ^2 L^2-
\frac{65 }{27} \pi ^2 N_l -\frac{461 }{54} \pi ^4 + 96 L^2 H_{2}(x_t)
   +96 L^2 H_{0,0}(x_t)
    \nonumber \\ &+96 L^2 H_{1,0}(x_t)+96 L^2
   H_{1,1}(x_t)+16 \pi ^2 L H_{0}(x_t)+16 \pi ^2 L H_{1}(x_t)-96 L
   H_{3}(x_t)
   \nonumber \\ & -96 L H_{1,2}(x_t)+96 L H_{2,0}(x_t)-288 L
   H_{2,1}(x_t)  +288 L H_{0,0,0}(x_t)+288 L H_{1,0,0}(x_t)
   \nonumber \\ & +96 L
   H_{1,1,0}(x_t)-288 L H_{1,1,1}(x_t)-16 \pi ^2 H_{2}(x_t)-288
   H_{4}(x_t)+16 \pi ^2 H_{0,0}(x_t)
     \nonumber \\ &+16 \pi ^2 H_{1,0}(x_t)
 -16 \pi ^2 H_{1,1}(x_t)-288 H_{1,3}(x_t)-96 H_{2,2}(x_t)
    -96 H_{3,0}(x_t)
    \nonumber \\ &+96 H_{3,1}(x_t)-96 H_{1,1,2}(x_t)-96
   H_{1,2,0}(x_t)+96 H_{1,2,1}(x_t)+96 H_{2,0,0}(x_t)
   -288 H_{2,1,0}(x_t)
   \nonumber \\ &+288 H_{2,1,1}(x_t)+288 H_{0,0,0,0}(x_t)
      +288
   H_{1,0,0,0}(x_t)+96 H_{1,1,0,0}(x_t)-288 H_{1,1,1,0}(x_t)
   \nonumber \\ &+288
   H_{1,1,1,1}(x_t)+1092 L \zeta (3)
+\frac{52 }{9} N_l \zeta (3)-\frac{286 }{3} \zeta(3) \, , \nonumber \\
\tilde{s}^{(2)}_{gg\, 12} &= \frac{134}{3} \pi ^2+\frac{52}{3} \pi ^2 L^2
   +\frac{4}{3} \pi ^2 L N_l-22 \pi ^2 L-\frac{20 }{9}\pi ^2 N_l-\frac{68 }{45}\pi ^4
 +\Biggl[104 L^3 +4 L^2 N_l -66 L^2 
 \nonumber \\ & -\frac{40}{3} L N_l+\frac{52}{3} \pi ^2 L +268
    L +\frac{4}{3} \pi ^2 N_l +\frac{112}{9} N_l+252 \zeta (3) -22 \pi ^2
-\frac{808}{3} \Biggr] H_{0}(x_t)
    \nonumber \\ & +\Biggl[ 104 L^3 +4 L^2
   N_l -66 L^2-\frac{40}{3} L N_l
+\frac{16}{3} \pi ^2 L +268 L +\frac{112}{9} N_l +\frac{4}{3} \pi ^2 N_l+324 \zeta (3) 
\nonumber \\ &-22 \pi ^2 -\frac{808}{3}\Biggr] H_{1}(x_t)+\Biggl[
 -104 L^2-8 L N_l +132 L+\frac{40}{3} N_l -\frac{16}{3} \pi ^2-268 \Biggr] H_{2}(x_t)
 \nonumber \\ &+ \Biggl[176 L^2 +8 L N_l -132 L -\frac{40}{3} N_l+\frac{52}{3} \pi ^2 +268
   \Biggr] H_{0,0}(x_t)
 +\Biggl[104 L^2+8 L N_l -132 L
 \nonumber \\ &-\frac{40}{3} N_l +\frac{16}{3} \pi ^2 +268 \Biggr] H_{1,0}(x_t)
+\Biggl[ -176 L^2-8 L N_l +132 L
+\frac{40}{3} N_l-\frac{52}{3} \pi ^2 
\nonumber \\ & -268 \Biggr] H_{1,1}(x_t)+\Bigl[ 72 L +8 N_l -132 \Bigr] H_{1,2}(x_t)
+\Bigl[ 72 L+8 N_l -132 \Bigr] H_{2,0}(x_t)+\Bigl[144 L
\nonumber \\ & +8 N_l-132\Bigr] H_{2,1}(x_t)
+\Bigl[ 216 L+8 N_l -132 \Bigr] H_{0,0,0}(x_t)
+\Bigl[ 144 L+8 N_l-132 \Bigr] H_{1,0,0}(x_t)
\nonumber \\ & 
   +\Bigl[ 216 L+8 N_l -132 \Bigr] H_{1,1,1}(x_t)
+\Bigl[ 8 N_l-132\Bigr] H_{3}(x_t)
+\Bigl[8 N_l-132 \Bigr] H_{1,1,0}(x_t)
      \nonumber \\ &+72 H_{4}(x_t)+72 H_{3,0}(x_t)-72 H_{1,1,2}(x_t)-144 H_{1,2,1}(x_t)+144
   H_{2,0,0}(x_t)
   \nonumber \\ & -72 H_{2,1,1}(x_t)+216 H_{0,0,0,0}(x_t)+72
   H_{1,0,0,0}(x_t)-72 H_{1,1,1,0}(x_t)-216 H_{1,1,1,1}(x_t) \nonumber \, ,\\
\tilde{s}^{(2)}_{gg\, 13} &= \frac{5 \pi ^4}{9} +\frac{20}{3} \pi ^2 L \Bigl[ H_{0}(x_t) + H_{1}(x_t)\Bigr]
   + \Biggl[ 40 L^2 -\frac{20}{3} \pi ^2 \Biggr] \Bigl[ H_{2}(x_t)  +H_{1,1}(x_t) \Bigr] 
\nonumber \\ & +\Biggl[ 40 L^2+\frac{20}{3} \pi ^2 \Biggr] \Bigl[ H_{0,0}(x_t) + H_{1,0}(x_t)\Bigr]
-40 L H_{3}(x_t) -40 L H_{1,2}(x_t) +40 L H_{2,0}(x_t)
\nonumber \\ & -120 L H_{2,1}(x_t)+120 L H_{0,0,0}(x_t)+120 L
   H_{1,0,0}(x_t)+40 L H_{1,1,0}(x_t)-120 L H_{1,1,1}(x_t)
   \nonumber \\ & -120 H_{4}(x_t) -120 H_{1,3}(x_t)-40
   H_{2,2}(x_t)-40 H_{3,0}(x_t)+40 H_{3,1}(x_t)-40
   H_{1,1,2}(x_t)
   \nonumber \\ & -40 H_{1,2,0}(x_t)+40 H_{1,2,1}(x_t)+40
   H_{2,0,0}(x_t) -120 H_{2,1,0}(x_t)+120 H_{2,1,1}(x_t) 
   \nonumber \\ & +120
   H_{0,0,0,0}(x_t)+120 H_{1,0,0,0}(x_t)+40 H_{1,1,0,0}(x_t)-120
   H_{1,1,1,0}(x_t)+120 H_{1,1,1,1}(x_t) \, ,\nonumber \\
\tilde{s}^{(2)}_{gg\, 22} &= \frac{15782}{27}-\frac{5252}{9} L+\frac{871 }{3}L^2-\frac{143}{3}L^3+ \frac{169}{3}  L^4
-\frac{2132}{81}N_l +\frac{728 }{27} N_l L-\frac{130 }{9}N_l L^2
 \nonumber \\ & +\frac{26}{9} N_l L^3
+\frac{2077 }{36}\pi ^2
 -\frac{33 }{2}\pi ^2 L+\frac{169}{9} \pi ^2 L^2
-\frac{155}{54} N_l \pi ^2+N_l \pi ^2 L-\frac{673 }{135}\pi ^4 
+\Biggl[ 78  L^3 
\nonumber \\ & +3 N_l  L^2-\frac{99}{2}  L^2-10 N_l  L+26 \pi ^2  L+201  L+N_l \pi ^2
-\frac{33}{2} \pi ^2 -202 +\frac{28}{3} N_l 
\nonumber \\ & +189 \zeta (3)\Biggr] H_{0}(x_t)
+\Biggl[ -78  L^3 -3 N_l  L^2+\frac{99}{2}  L^2+10 N_l  L-201  L-\frac{28}{3}
   N_l +N_l \pi ^2 
   \nonumber \\ & -\frac{33}{2} \pi ^2 +202 -135  \zeta (3)\Biggr]H_{1}(x_t)
+\Bigl[ 102 L^2+6 N_l L-99 L-10 N_l+201 \Bigr] H_{2}(x_t)
\nonumber \\ & 
+\Bigl[ 210 L^2+6 N_l L-99 L-10 N_l+26 \pi ^2+201 \Bigr] H_{0,0}(x_t)
+\Bigl[ 102 L^2+6 N_l L-99 L
\nonumber \\ &
-10 N_l +17 \pi ^2 +201 \Bigr] H_{1,0}(x_t)
+\Bigl[ 210 L^2+6 N_l L-99 L-10 N_l -9 \pi ^2 +201 \Bigr] H_{1,1}(x_t)
\nonumber \\ &
+\Bigl[ -24 L-6 N_l+99 \Bigr] H_{3}(x_t)
+\Bigl[ -132 L-6 N_l+99 \Bigr] H_{1,2}(x_t)
+\Bigl[ -288 L-6 N_l
\nonumber \\ &
+99\Bigr] H_{2,1}(x_t)
+\Bigl[132 L+6 N_l -99\Bigr] H_{2,0}(x_t)
+\Bigl[396 L+6 N_l -99\Bigr] H_{0,0,0}(x_t)
\nonumber \\ & 
+\Bigl[288 L+6 N_l -99\Bigr] H_{1,0,0}(x_t)
+\Bigl[24 L+6 N_l -99\Bigr] H_{1,1,0}(x_t)
+\Bigl[-396 L-6 N_l 
\nonumber \\ &
+99\Bigr] H_{1,1,1}(x_t)-126 H_{4}(x_t)-72 H_{1,3}(x_t)+30 H_{2,2}(x_t) +84 H_{3,0}(x_t)+132 H_{3,1}(x_t)
      \nonumber \\ &+84 H_{1,1,2}(x_t)+30 H_{1,2,0}(x_t)+240
   H_{1,2,1}(x_t)+240 H_{2,0,0}(x_t)-72 H_{2,1,0}(x_t)+288
   H_{2,1,1}(x_t)
   \nonumber \\ & +396 H_{0,0,0,0}(x_t)+288 H_{1,0,0,0}(x_t) +132 H_{1,1,0,0}(x_t)-126 H_{1,1,1,0}(x_t)+396 H_{1,1,1,1}(x_t)
      \nonumber \\ &-\frac{143 }{3} \zeta(3) +\frac{26}{9}  N_l \zeta (3) +546 \zeta (3) L \, , \nonumber \\
\tilde{s}^{(2)}_{gg\, 23} &= \frac{335}{18} \pi ^2-\frac{55}{6}  \pi ^2 L+\frac{65 }{9}\pi ^2 L^2-\frac{25}{27}  N_l \pi ^2
+\frac{5}{9} N_l \pi ^2 L-\frac{23}{108} \pi ^4
+\Biggl[ \frac{130}{3} L^3+\frac{5}{3} N_l L^2-\frac{55}{2} L^2
\nonumber \\ & 
-\frac{50}{9} N_l  L
+\frac{110}{9} \pi ^2 L+\frac{335}{3}  L+\frac{140}{27} N_l 
+\frac{5}{9} N_l \pi ^2 -\frac{55}{6} \pi ^2 -\frac{1010}{9}+105\zeta (3)\Biggr] H_{0}(x_t) 
\nonumber \\ &
+\Biggl[\frac{130}{3}  L^3
+\frac{5}{3} N_l  L^2-\frac{55}{2}  L^2-\frac{50}{9} N_l L+\frac{20}{9} \pi ^2 
 L+\frac{335}{3}  L+\frac{140}{27} N_l +\frac{5}{9} N_l \pi ^2 -\frac{55}{6} \pi ^2 -\frac{1010}{9} 
\nonumber \\ &
 +135 \zeta (3)\Biggr] H_{1}(x_t)
+\Biggl[ -\frac{130}{3} L^2-\frac{10}{3} N_l L+55 L+\frac{50}{9} N_l -\frac{20}{9} \pi ^2-\frac{335}{3} \Biggr] H_{2}(x_t)
+\Biggl[\frac{310}{3} L^2
\nonumber \\ &
+\frac{10}{3} N_l L-55 L+\frac{110}{9}\pi ^2+\frac{335}{3} -\frac{50}{9} N_l \Biggr] H_{0,0}(x_t)
+ \Biggl[\frac{130}{3} L^2-55 L+\frac{10}{3} N_l L-\frac{50}{9} N_l 
\nonumber \\ &
+\frac{65}{9} \pi ^2+\frac{335}{3} \Biggr] H_{1,0}(x_t)
+\Biggl[ -\frac{310}{3} L^2-\frac{10}{3} N_l L+55 L+\frac{50}{9} N_l-\frac{65}{9} \pi ^2-\frac{335}{3} \Biggr] H_{1,1}(x_t)
\nonumber \\ &
+\Biggl[60 L+\frac{10}{3} N_l -55\Biggr] \Bigl[H_{1,2}(x_t) + H_{2,0}(x_t)\Bigr]
+\Biggl[120 L+\frac{10}{3} N_l-55 \Biggr]H_{2,1}(x_t)
+\Biggl[180 L
\nonumber \\ &
+\frac{10}{3} N_l-55\Biggr] \Bigl[H_{0,0,0}(x_t) + H_{1,1,1}(x_t)\Bigr]
+\Biggl[120 L+\frac{10}{3} N_l -55 \Biggr] H_{1,0,0}(x_t)
+\Biggl[ \frac{10}{3} N_l 
\nonumber \\ & 
-55 \Biggr] H_{3}(x_t)+30 H_{4}(x_t)
-30 H_{2,2}(x_t)+60 H_{3,0}(x_t)-60 H_{3,1}(x_t)+\frac{10}{3} N_l H_{1,1,0}(x_t)
\nonumber \\ &
-55 H_{1,1,0}(x_t) -60 H_{1,1,2}(x_t)+30 H_{1,2,0}(x_t)-120 H_{1,2,1}(x_t)+120
   H_{2,0,0}(x_t)
\nonumber \\ &
   -120 H_{2,1,1}(x_t)+180 H_{0,0,0,0}(x_t)+120
   H_{1,0,0,0}(x_t)+60 H_{1,1,0,0}(x_t)
     -30 H_{1,1,1,0}(x_t)
     \nonumber \\ &
     -180 H_{1,1,1,1}(x_t) \, .\nonumber \\
\tilde{s}^{(2)}_{gg\, 33} &=\frac{78910}{243}-\frac{26260}{81} L+\frac{4355}{27} L^2-\frac{715}{27} L^3+ \frac{845}{27} L^4
-\frac{10660}{729}  N_l+\frac{3640}{243}N_l L-\frac{650}{81} N_l L^2
\nonumber \\ &
+\frac{130}{81} N_l L^3 +\frac{10385}{324} \pi ^2-\frac{55}{6} \pi ^2 L+\frac{845}{81} \pi ^2 L^2
-\frac{775}{486} N_l \pi^2+\frac{5}{9} N_l \pi ^2 L
-\frac{763}{243} \pi ^4
+\Biggl[\frac{130}{3} L^3
\nonumber \\ &
+\frac{5}{3} N_lL^2-\frac{55}{2} L^2+\frac{335}{3} L-\frac{50}{9} N_l  L
+10 \pi ^2 L +\frac{5}{9} N_l \pi ^2
   -\frac{55}{6} \pi ^2 -\frac{1010}{9}+\frac{140}{27} N_l 
      \nonumber \\ &+105  \zeta (3)
   \Biggr] H_{0}(x_t)  
+\Biggl[-\frac{130}{3}L^3-\frac{5}{3} N_l  L^2+\frac{55}{2}  L^2+\frac{50}{9} N_l  L-\frac{40}{9} \pi ^2
    L-\frac{335}{3}  L+\frac{5}{9} N_l \pi ^2  
            \nonumber \\ &-\frac{55}{6} \pi ^2 +\frac{1010}{9} -\frac{140}{27} N_l -75
   \zeta (3)\Biggr] H_{1}(x_t)
 +\Biggl[30 L^2+\frac{10}{3} N_l  L-55  L-\frac{50}{9} N_l +\frac{40}{9} \pi ^2
 \nonumber \\ &
 +\frac{335}{3}\Biggr] H_{2}(x_t)
+\Biggl[90  L^2+\frac{10}{3} N_l  L-55  L-\frac{50}{9} N_l +10 \pi ^2 +\frac{335}{3} \Biggr] H_{0,0}(x_t)
+\Biggl[30  L^2
\nonumber \\  &
+\frac{10}{3} N_l  L-55  L-\frac{50}{9} N_l +5 \pi ^2 +\frac{335}{3} \Biggr] H_{1,0}(x_t)
+\Biggl[ 90  L^2+\frac{10}{3} N_l  L-55  L-\frac{50}{9} N_l -\frac{5}{9} \pi ^2 
\nonumber \\ &+\frac{335}{3} \Biggr] H_{1,1}(x_t)
+\Biggl[\frac{40}{3} L-\frac{10}{3} N_l +55 \Biggr] H_{3}(x_t)
+\Biggl[ \frac{140}{3}  L+\frac{10}{3} N_l -55\Biggr] \Bigl[ H_{2,0}(x_t)  
\nonumber \\ &
- H_{1,2}(x_t)\Bigr]
+\Biggl[ -80 L-\frac{10}{3} N_l +55 \Biggr] H_{2,1}(x_t)
+\Biggl[ 140 L+\frac{10}{3} N_l -55 \Biggr] \Bigl[H_{0,0,0}(x_t) 
\nonumber \\ &
- H_{1,1,1}(x_t)\Bigr]
+\Biggl[ 80 L+\frac{10}{3} N_l -55\Biggr] H_{1,0,0}(x_t) +
\Biggl[-\frac{40}{3}  L-55 \Biggr] H_{1,1,0}(x_t)
+10 H_{4}(x_t)
      \nonumber \\ &+40
   H_{1,3}(x_t)+\frac{130}{3} H_{2,2}(x_t)
      +\frac{220}{3}
   H_{3,0}(x_t)+\frac{140}{3} H_{3,1}(x_t)+\frac{10}{3} N_l H_{1,1,0}(x_t)
      \nonumber \\ &+\frac{220}{3} H_{1,1,2}(x_t)+\frac{130}{3}
   H_{1,2,0}(x_t)+\frac{320}{3} H_{1,2,1}(x_t)
   +\frac{320}{3} H_{2,0,0}(x_t)+40
   H_{2,1,0}(x_t)
      \nonumber \\ &
   +80 H_{2,1,1}(x_t)+140 H_{0,0,0,0}(x_t)+80
   H_{1,0,0,0}(x_t)+\frac{140}{3} H_{1,1,0,0}(x_t)
      +10 H_{1,1,1,0}(x_t)
      \nonumber \\ &
      +140 H_{1,1,1,1}(x_t)+\frac{130 N_l \zeta (3)}{81}-\frac{715 }{27}\zeta (3) +\frac{910 L}{3}\zeta (3) \, .
\end{align}


\begin{thebibliography}{99}

\bibitem{Mangano:1991jk} 
  M.~L.~Mangano, P.~Nason and G.~Ridolfi,
  Nucl.\ Phys.\ B {\bf 373}, 295 (1992).

\bibitem{Kidonakis:1997gm} 
  N.~Kidonakis and G.~F.~Sterman,
  Nucl.\ Phys.\ B {\bf 505}, 321 (1997)
  [hep-ph/9705234].

\bibitem{Almeida:2008ug} 
  L.~G.~Almeida, G.~F.~Sterman and W.~Vogelsang,
  Phys.\ Rev.\ D {\bf 78}, 014008 (2008)
  [arXiv:0805.1885 [hep-ph]].

\bibitem{Ahrens:2010zv} 
  V.~Ahrens, A.~Ferroglia, M.~Neubert, B.~D.~Pecjak and L.~L.~Yang,
  JHEP {\bf 1009}, 097 (2010)
  [arXiv:1003.5827 [hep-ph]].

\bibitem{Appell:1988ie} 
  D.~Appell, G.~F.~Sterman and P.~B.~Mackenzie,
  Nucl.\ Phys.\ B {\bf 309}, 259 (1988).

\bibitem{Catani:1998tm} 
  S.~Catani, M.~L.~Mangano and P.~Nason,
  JHEP {\bf 9807}, 024 (1998)
  [hep-ph/9806484].

\bibitem{Becher:2007ty} 
  T.~Becher, M.~Neubert and G.~Xu,
  JHEP {\bf 0807}, 030 (2008)
  [arXiv:0710.0680 [hep-ph]].

\bibitem{Bauer:2010jv} 
  C.~W.~Bauer, N.~D.~Dunn and A.~Hornig,
  arXiv:1010.0243 [hep-ph].

\bibitem{Ahrens:2009uz} 
  V.~Ahrens, A.~Ferroglia, M.~Neubert, B.~D.~Pecjak and L.~L.~Yang,
  Phys.\ Lett.\ B {\bf 687}, 331 (2010)
  [arXiv:0912.3375 [hep-ph]].

\bibitem{Baernreuther:2012ws} 
  P.~Baernreuther, M.~Czakon and A.~Mitov,
  arXiv:1204.5201 [hep-ph].

\bibitem{Czakon:2012zr} 
  M.~Czakon and A.~Mitov,
  arXiv:1207.0236 [hep-ph].

\bibitem{Ferroglia:2012ku} 
  A.~Ferroglia, B.~D.~Pecjak and L.~L.~Yang,
  arXiv:1205.3662 [hep-ph].

\bibitem{Melnikov:2004bm} 
  K.~Melnikov and A.~Mitov,
  Phys.\ Rev.\ D {\bf 70}, 034027 (2004)
  [hep-ph/0404143].

\bibitem{Anastasiou:2000kg} 
  C.~Anastasiou, E.~W.~N.~Glover, C.~Oleari and M.~E.~Tejeda-Yeomans,
  Nucl.\ Phys.\ B {\bf 601}, 318 (2001)
  [hep-ph/0010212].

\bibitem{Anastasiou:2000mv} 
  C.~Anastasiou, E.~W.~N.~Glover, C.~Oleari and M.~E.~Tejeda-Yeomans,
  Phys.\ Lett.\ B {\bf 506}, 59 (2001)
  [hep-ph/0012007].

\bibitem{Glover:2001af} 
  E.~W.~N.~Glover, C.~Oleari and M.~E.~Tejeda-Yeomans,
  Nucl.\ Phys.\ B {\bf 605}, 467 (2001)
  [hep-ph/0102201].

\bibitem{Glover:2001rd} 
  E.~W.~N.~Glover and M.~E.~Tejeda-Yeomans,
  JHEP {\bf 0105}, 010 (2001)
  [hep-ph/0104178].

\bibitem{Anastasiou:2001sv} 
  C.~Anastasiou, E.~W.~N.~Glover, C.~Oleari and M.~E.~Tejeda-Yeomans,
  Nucl.\ Phys.\ B {\bf 605}, 486 (2001)
  [hep-ph/0101304].

\bibitem{Belitsky:1998tc} 
  A.~V.~Belitsky,
  Phys.\ Lett.\ B {\bf 442}, 307 (1998)
  [hep-ph/9808389].

\bibitem{Becher:2005pd} 
  T.~Becher and M.~Neubert,
  Phys.\ Lett.\ B {\bf 633}, 739 (2006)
  [hep-ph/0512208].

\bibitem{Kelley:2011ng} 
  R.~Kelley, M.~D.~Schwartz, R.~M.~Schabinger and H.~X.~Zhu,
  Phys.\ Rev.\ D {\bf 84}, 045022 (2011)
  [arXiv:1105.3676 [hep-ph]].

\bibitem{Monni:2011gb} 
  P.~F.~Monni, T.~Gehrmann and G.~Luisoni,
  JHEP {\bf 1108}, 010 (2011)
  [arXiv:1105.4560 [hep-ph]].

\bibitem{Hornig:2011iu} 
  A.~Hornig, C.~Lee, I.~W.~Stewart, J.~R.~Walsh and S.~Zuberi,
  JHEP {\bf 1108}, 054 (2011)
  [arXiv:1105.4628 [hep-ph]].

\bibitem{Li:2011zp} 
  Y.~Li, S.~Mantry and F.~Petriello,
  Phys.\ Rev.\ D {\bf 84}, 094014 (2011)
  [arXiv:1105.5171 [hep-ph]].

\bibitem{Kelley:2011aa} 
  R.~Kelley, M.~D.~Schwartz, R.~M.~Schabinger and H.~X.~Zhu,
  arXiv:1112.3343 [hep-ph].

\bibitem{Becher:2012za} 
  T.~Becher, G.~Bell and S.~Marti,
  JHEP {\bf 1204}, 034 (2012)
  [arXiv:1201.5572 [hep-ph]].

\bibitem{Gatheral:1983cz} 
  J.~G.~M.~Gatheral,
  Phys.\ Lett.\ B {\bf 133}, 90 (1983).

\bibitem{Frenkel:1984pz} 
  J.~Frenkel and J.~C.~Taylor,
  Nucl.\ Phys.\ B {\bf 246}, 231 (1984).

\bibitem{Aybat:2006wq} 
  S.~M.~Aybat, L.~J.~Dixon and G.~F.~Sterman,
  Phys.\ Rev.\ Lett.\  {\bf 97}, 072001 (2006)
  [hep-ph/0606254].

\bibitem{Aybat:2006mz} 
  S.~M.~Aybat, L.~J.~Dixon and G.~F.~Sterman,
  Phys.\ Rev.\ D {\bf 74}, 074004 (2006)
  [hep-ph/0607309].

\bibitem{Becher:2009cu} 
  T.~Becher and M.~Neubert,
  Phys.\ Rev.\ Lett.\  {\bf 102}, 162001 (2009)
  [arXiv:0901.0722 [hep-ph]].

\bibitem{Gardi:2009qi} 
  E.~Gardi and L.~Magnea,
  JHEP {\bf 0903}, 079 (2009)
  [arXiv:0901.1091 [hep-ph]].

\bibitem{Becher:2009qa} 
  T.~Becher and M.~Neubert,
  JHEP {\bf 0906}, 081 (2009)
  [arXiv:0903.1126 [hep-ph]].

\bibitem{Catani:1996jh} 
  S.~Catani and M.~H.~Seymour,
  Phys.\ Lett.\ B {\bf 378}, 287 (1996)
  [hep-ph/9602277].

\bibitem{Catani:1996vz} 
  S.~Catani and M.~H.~Seymour,
  Nucl.\ Phys.\ B {\bf 485}, 291 (1997)
  [Erratum-ibid.\ B {\bf 510}, 503 (1998)]
  [hep-ph/9605323].

\bibitem{vanNeerven:1985xr} 
  W.~L.~van Neerven,
  Nucl.\ Phys.\ B {\bf 268}, 453 (1986).

\bibitem{Remiddi:1999ew} 
  E.~Remiddi and J.~A.~M.~Vermaseren,
  Int.\ J.\ Mod.\ Phys.\ A {\bf 15}, 725 (2000)
  [hep-ph/9905237].

\bibitem{Huber:2005yg} 
  T.~Huber and D.~Maitre,
  Comput.\ Phys.\ Commun.\  {\bf 175}, 122 (2006)
  [hep-ph/0507094].

\bibitem{Maitre:2005uu} 
  D.~Maitre,
  Comput.\ Phys.\ Commun.\  {\bf 174}, 222 (2006)
  [hep-ph/0507152].

\bibitem{Ferroglia:2009ep} 
  A.~Ferroglia, M.~Neubert, B.~D.~Pecjak and L.~L.~Yang,
  Phys.\ Rev.\ Lett.\  {\bf 103}, 201601 (2009)
  [arXiv:0907.4791 [hep-ph]].

\bibitem{Ferroglia:2009ii} 
  A.~Ferroglia, M.~Neubert, B.~D.~Pecjak and L.~L.~Yang,
  JHEP {\bf 0911}, 062 (2009)
  [arXiv:0908.3676 [hep-ph]].

\bibitem{Kidonakis:2010dk} 
  N.~Kidonakis,
  Phys.\ Rev.\ D {\bf 82}, 114030 (2010)
  [arXiv:1009.4935 [hep-ph]].

\bibitem{Ahrens:2011mw} 
  V.~Ahrens, A.~Ferroglia, M.~Neubert, B.~D.~Pecjak and L.~-L.~Yang,
  JHEP {\bf 1109}, 070 (2011)
  [arXiv:1103.0550 [hep-ph]].

\end{thebibliography}
\end{document}